\documentclass{aastex61}
\usepackage[utf8]{inputenc}
\usepackage{amsfonts,amsmath, amssymb, bm}
\usepackage{fontenc}
\usepackage{graphicx}

\newcommand{\abs}[1]{\left| #1 \right|}

\begin{document}

\title{Reconnection fluxes in eruptive and confined flares and implications for superflares on the Sun}
\author[0000-0002-2004-222X]{Johannes Tschernitz}
\affiliation{Institute of Physics, University of Graz, Universit\"atsplatz 5, 8010 Graz, Austria}
\email{johannes.tschernitz@edu.uni-graz.at}
\correspondingauthor{Astrid  M. Veronig}
\email{astrid.veronig@uni-graz.at}

\author[0000-0003-2073-002X]{Astrid M. Veronig}
\affiliation{Institute of Physics, University of Graz, Universit\"atsplatz 5, 8010 Graz, Austria}
\affiliation{Kanzelh\"ohe Observatory for Solar and Environmental Research, University of Graz, Kanzelh\"ohe 19, A-9521 Kanzelh\"ohe, Austria}
\email{astrid.veronig@uni-graz.at}
\author[0000-0001-8985-2549]{Julia K. Thalmann}
\affiliation{Institute of Physics, University of Graz, Universit\"atsplatz 5, 8010 Graz, Austria}
\email{julia.thalmann@uni-graz.at}
\author{J\"urgen Hinterreiter}
\affiliation{Institute of Physics, University of Graz, Universit\"atsplatz 5, 8010 Graz, Austria}
\email{juergen.hinterreiter@edu.uni-graz.at}
\author[0000-0003-4811-0543]{Werner P\"otzi}
\affiliation{Kanzelh\"ohe Observatory for Solar and Environmental Research, University of Graz, Kanzelh\"ohe 19, A-9521 Kanzelh\"ohe, Austria}
\email{werner.poetzi@uni-graz.at}

\begin{abstract}
We study the energy release process of a set of 51 flares (32 confined, 19 eruptive) ranging from GOES class B3 to X17. We use H$\alpha$ filtergrams from Kanzelh\"ohe Observatory together with SDO HMI and SOHO MDI magnetograms to derive magnetic reconnection fluxes and rates. The flare reconnection flux is strongly correlated with the peak of the GOES 1-8 \AA\ soft X-ray flux (c=0.92, in log-log space), both for confined and eruptive flares.
Confined flares of a certain GOES class exhibit smaller ribbon areas but larger magnetic flux densities  in the flare ribbons (by a factor of 2). 
 In the largest events, up to $\approx$50\%\ of the magnetic flux of the active region (AR) causing the flare is involved in the flare magnetic reconnection. These findings allow us 
 to extrapolate toward the largest solar flares possible.
A complex solar AR hosting a magnetic flux of $2\cdot 10^{23}\, \mathrm{Mx}$, which is in line with the largest AR fluxes directly measured, is capable of producing an X80 flare, which corresponds to a bolometric energy of about $7 \cdot 10^{32}$ ergs. Using a magnetic flux estimate of  $6\cdot 10^{23}\, \mathrm{Mx}$  for the largest solar AR observed, we find that flares of GOES class $\approx$X500 could be produced ($E_{\rm bol} \approx 3 \cdot 10^{33}$ ergs). These estimates suggest that the present day's Sun is capable of producing flares and related space weather events that may be more than an order of magnitude stronger than have been observed to date.
\end{abstract}


\section{Introduction}
During solar flares huge amounts of energy are released over a course of tens of minutes \citep{Fletcher2011}. The energy is stored in stressed coronal magnetic fields and released via magnetic reconnection \citep{Priest2002}. While a flare can occur together with a coronal mass ejection (CME), flares without CMEs or CMEs without flares are also frequently observed. \cite{Yashiro2005} studied the association rate of flares and CMEs. In their study, they found an association rate of 20\%\ for small (C-class) flares, which steeply increases with increasing flare strength. About 50\%\ of the M-class flares are associated with a CME while the percentage increases up to \textgreater 90\%\ for X-class flares. Finally, all flares $\gtrsim$X5 are associated with CMEs.

The widely accepted model for the physical processes in eruptive two-ribbon flares is the CSHKP model, based on the works by \cite{Carmichael1964}, \cite{Sturrock1966}, \cite{Hirayama1974} and \cite{Kopp1976}. A rising filament leads to the formation of a current sheet below it, and due to some instabilities magnetic reconnection sets in. Within this framework, and its extension to three dimensions, many of the observed flare features can be successfully explained \citep[cf. review by][]{Janvier2015}. The flare ribbon evolution gives indirect evidence of the magnetic reconnection process and can be observed at H$\alpha$, (E)UV and hard X-ray (HXR) wavelengths \citep[cf. reviews by][]{Fletcher2011,Holman2016,Benz2017}. The flare ribbons mark the chromospheric footpoints of newly reconnected magnetic field and are caused by energetic particles transported downwards along the field lines, in particular by accelerated electron beams \citep[for reviews on the underlying reconnection process see, e.g.,][]{Priest2002,Shibata2011}.

The reconnection process itself takes place in the corona, and is in general not accessible to direct observations. Therefore, we have to rely on other means to get insight into the associated energy release processes. \cite{Forbes1984} showed that in 2D the reconnection rate in a flare is directly proportional to the reconnection electric field in the corona, $E_{0}$, and can be derived from observed quantities as
 \begin{equation}
 E_{0}=B_{n}V_{R},
 \label{eq:vxb}
 \end{equation}
with $B_{n}$ the component of the photospheric magnetic field normal to the surface and $V_{R}$ the separation velocity of the flare ribbons perpendicular to the polarity inversion line. \cite{Forbes2000} generalized Eq.~\ref{eq:vxb} to three dimensions to overcome the limitations of a two dimensional magnetic field configuration required in Eq.~\ref{eq:vxb}. They considered the magnetic flux, $\varphi$, of one polarity swept by the flare ribbons:
\begin{equation}
 \varphi=\iint_{A}\mathbf{B}\cdot d\mathbf{a}.
 \label{eq:flux}
\end{equation}
By taking the time derivative of Eq.~\ref{eq:flux}, one obtains an expression for the rate of change in the reconnection magnetic flux over time in the form 
\begin{equation}
 \dot{\varphi}=\int_{C}\left(\mathbf{B}\times\mathbf{V}_{R}\right)\cdot d\mathbf{l}=\oint_{C}\mathbf{E}_0\cdot d\mathbf{l},
 \label{eq:phidot}
\end{equation}
with $\mathbf{E_{0}}$ as the electric field along a separator line, $\mathbf{B}$ the magnetic field vector measured in the photosphere and $C$ the closed curve surrounding the newly closed area $A$. A separator line is a field line connecting a pair of magnetic neutral points. Due to line tying, in Eq.~\ref{eq:phidot} the term $\partial\mathbf{B}/\partial t$ is neglected. Eq.~\ref{eq:phidot} gives the voltage drop along the separator line and corresponds to the rate of open flux converted to closed flux \citep{Forbes2000}. 

This method was used in several case studies to derive magnetic reconnection rates in flares \citep[e.g.,][]{Qiu2005,Miklenic2009,Veronig2015}. \cite{Qiu2005} showed that the speed of the associated CME correlates with the amount of flux processed during magnetic reconnection. They also noted that the amount of flux in opposite polarity regions involved in the reconnection should equal each other, and that this balance is never perfectly obtained due to measurement uncertainties. Thus they argue that a ratio of positive and negative magnetic flux in the range [0.5, 2] is a reasonable range to identify flux balance. \cite{Veronig2015} found a correlation between the total reconnection flux and the peak soft X-ray (SXR) flux of $c=0.78$, using data of 27 eruptive flares collected from 4 different studies. In addition, it was found that $\dot{\varphi}\left(t\right)$  resembles the flare HXR time profile and the SXR time derivative, both proxies of the energy release rate in the flare \citep{Miklenic2009,Veronig2015}. 

This empirical relationship between the HXR (and microwave) flux produced by flare-accelerated electrons and the time derivative of the SXR flux was termed 'Neupert effect' by \cite{Hudson1991a}, based on its first recognition in the paper by \cite{Neupert1968}. Therafter, it has been established in several observational studies \citep[e.g.][]{Dennis1993,Veronig2002,Veronig2005}. The HXR flux in solar flares is due to nonthermal bremsstrahlung from electrons that are accelerated during the energy release process by magnetic reconnection in the corona, and precipitate downward along the newly reconnected (closed) field lines. Most of the nonthermal HXR emission is produced when the electrons reach the chromosphere, due to the steep gradient in the column density, emitting nonthermal bremsstrahlung when scattering off the ions of the ambient plasma \citep{Brown1971,Brown1973}. Thus, the HXR flux is an instantaneous response to this energy input by flare-accelerated electrons, which carry a large fraction of the total energy released during a flare \citep[e.g.][]{ Hudson1991a,Dennis2003}. 
The increase of the soft X-ray (SXR) flux during a flare is due to the response of the chromosphere and corona to the heating by collisions of the high-energy flare electrons with the thermal electrons of the ambient plasma. The strong heating of the chromosphere causes the upper chromospheric layers to convect into the corona, in a process called 'chromospheric evaporation' \citep{Neupert1968,Fisher1985}. This process increases the density and the temperature in the flare loops, which results in an continuous increase of the SXR emission during the flare impulsive phase. Thus, the SXR flux profile is an indicator of the cumulated flare energy (ignoring cooling effects to first order), whereas the HXRs are a proxy of the instantaneous flare energy release rate \citep{Veronig2005}.

Recently, several statistical studies of flare reconnection fluxes and their relation to the flare energy in terms of the GOES SXR peak flux  became available.
 \cite{Toriumi2017} studied all flare events of GOES class $\ge$M5.0, which occurred in the period May 2010 to April 2016 and within 45$^\circ$\ from the center of the solar disk, in total a number of 51 events. They found a low correlation between the GOES peak flux and the total magnetic flux swept by flare ribbons derived from SDO AIA 1600 \AA\ data, c=0.37 in log-log space. \cite{Kazachenko2017} analyzed more than 3000 flares (GOES class C1.0 and larger) that occurred between May 2010 to April 2016 and within 45$^\circ$\ from disk center, also using SDO AIA 1600 \AA\ data. They found a Spearman ranking correlation coefficient of 0.66 between the peak SXR flux and the magnetic flux swept by the flare ribbons. In both studies the largest flare in the data set was of GOES class X5.4.

In the present study we perform a similar analysis, but with the following important differences. 1) Our data set comprises a large range from B3 to X17, i.e. covering almost four orders of magnitude in GOES class. 2) We use full-disk H$\alpha$\ filtergrams for the analysis of the flare-associated ribbons and corresponding estimation of the reconnection fluxes. 3) We study the relationship between the reconnection flux and the flare GOES class separately for confined and eruptive flares. 4) We discuss the implications of our findings with respect to the largest flares that can be expected from the present day's Sun, and their relation to stellar superflares 
on solar-like stars recently discovered by the Kepler satellite \citep{Maehara2012}.

\section{Data set}
\label{sec:data}

In this study we use high cadence full disk H$\alpha$\ filtergrams and line-of-sight (LOS) magnetograms to calculate the magnetic reconnection fluxes and magnetic reconnection rates of a homogeneous data set of 51 confined and eruptive flares, distributed over four orders in GOES class (cf. Table \ref{tab:dist}). High-cadence H$\alpha$\ filtergrams are provided by Kanzelh\"ohe Observatory for Solar and Environmental Research (KSO; \url{http://kso.ac.at}). We selected the events based on the flares observed by KSO\footnote{A list of flares observed at KSO since 1948 can be found at \url{http://cesar.kso.ac.at/flare_data/kh_flares_query.php}}. 

For our data set we aimed for a broad distribution over the flare classes and to include eruptive as well as confined flares of different classes. We limited the selection to events within 45$^\circ$\ of the center of the solar disk in order to reduce uncertainties due to projection effects. Between the years 2000 and 2015, KSO observed a total of 4357 flares, out of which 15 were of 
H$\alpha$\ importance class 3 or 4. All of them occurred close enough to the center of the solar disc and therefore are selected for our study. Then we selected an appropriate number of well observed smaller flares, beginning with class 2 flares. Finally we selected a number of suitable flares of classes 1 and subflares (S), beginning in 2014 and going backwards. While compiling the data set, we put emphasis on including also large confined flares while maintaining a good distribution among all flare classes. The resulting data set contains a total of 51 flares distributed over all flare classes (cf. Tables \ref{tab:dist} and \ref{tab:events}). Note that in an accompanying paper the same data set was used to study the flare ribbon characteristics (initial ribbon distance, separation and speed) and the reconnection electric field (Hinterreiter et al. 2017).

The H$\alpha$\ filtergrams for two of the flares were provided by other observatories, allowing us to include also the strongest flares in solar cycle 23 that occurred close to the disk center. For the X17 flare on 2003 October 28 we use data from Udaipur Solar Observatory (USO), and for the X10 flare on 2003 October 29 we use data from the Improved Solar Observing Optical Network (ISOON). KSO observations for the X17 flare on 2003 October 28 have large data gaps during the impulsive phase and are therefore not suitable for our analysis, while the X10 flare on 2003 October 29 occurred at KSO night. 

For each event, our analysis is based on a time series of H$\alpha$\ filtergrams starting 5 minutes prior to the nominal flare start time and ending 5 minutes after the nominal end time. The respective times and can be found in Table \ref{tab:events} and are taken from the list of observed flares from KSO. For the event on 2003 October 29 the times provided by NOAA's National Geophysical Data Center (NGDC)\footnote{\url{ftp://ftp.ngdc.noaa.gov/STP/space-weather/solar-data/solar-features/solar-flares/h-alpha/reports/}} are used.  

\begin{table}
\centering
\tabletypesize{\footnotesize}
\renewcommand\arraystretch{0.9}
\caption{Distribution of the events with respect to the GOES SXR flare classes.}
\label{tab:dist}
\begin{tabular}{cccc}
\hline \hline
GOES-class & N$_{\rm confined}$ & N$_{\rm eruptive}$ & N$_{\rm total}$\\ \hline
B & 7 & 0 & 7 \\
C & 12 & 3 & 15 \\
M & 7 & 11 & 18 \\
X & 6 & 3 & 9 \\
$\ge$X10 &  0 & 2 & 2 \\ \hline
Total & 32 & 19 & 51 \\ \hline
\end{tabular}
\end{table}

The observations at KSO are performed with a refractor with an aperture number of d/f=100/2000. The filter is a Lyot band-pass filter centered on the H$\alpha$\ spectral line with a full width at half maximum (FWHM) of 0.07 nm. An interference filter with a FWHM of 10 nm is placed in the light path for thermal protection \citep{Poetzi2015}. Since the year 2000, three different CCD cameras were used at KSO. From 2000 on regular observations were carried out with a 1k~$\times$~1k 8-bit CCD with a spatial resolution of 2.2\arcsec. The cadence was 1 image per minute \citep{Otruba1999}. The CCD was replaced with a 10 bit CCD with the same number of pixels in mid 2005. The observing cadence stayed the same \citep{Otruba2003}. 
Since the year 2008, the observations are carried out with a 2048~$\times$~2048 pixels 12 bit CCD with a spatial resolution of approximately 1\arcsec\ per pixel, and an increased observing cadence of 6 seconds  \citep{Poetzi2015}. 

The USO high resolution H$\alpha$\ filtergrams are obtained by a 15-cm aperture f/15 telescope and a 12-bit CCD. The images have a cadence of approximately 30 seconds. Comparison and co-alignment with KSO data, which partially covered this event, resulted in a plate scale of $\sim$0.6\arcsec. 
The ISOON data are obtained by a 12-bit CCD with a plate scale of $\sim$1{\arcsec} and a  cadence of 1 minute \citep{Neidig1998}.

The H$\alpha$\ data set is complemented by either 96m full disk line-of-sight (LOS) magnetograms from the \textit{Michelson Doppler Imager} aboard the \textit{Solar and Heliospheric Observatory} (SOHO/MDI) or 720s full disk LOS magnetograms from the  \textit{Helioseismic and Magnetic Imager} aboard the \textit{Solar Dynamics Observatory} (SDO/HMI) for flares after 2010 May 1. The MDI magnetograms are recorded with a CCD with 1024 $\times$ 1024 pixels and have a spatial resolution of 4{\arcsec} \citep{Scherrer1995}. HMI provides 4096 $\times$ 4096 pixel LOS magnetograms with a spatial resolution of about 1{\arcsec} \citep{Schou2012}. We use the 720s LOS magnetograms from HMI because of the better signal-to-noise ratio compared to the 45s LOS magnetograms \citep{Couvidat2016}. For each event we selected the latest magnetogram available before the start of the flare as listed in Table \ref{tab:events}. 

The CME association for the flares under study was determined using the SOHO/LASCO CME catalog \footnote{\url{https://cdaw.gsfc.nasa.gov/CME_list/}}. 
If there were CMEs with back-extrapolated onset times within an hour of the flare start time, then these events were considered as candidates for eruptive flares. In addition, we verified the CME-association of the flares under study by visual comparison of LASCO-C2 observations to the flare observations, demanding that the CME position angle coincides with the quadrant of the flare location on the solar disk.
For each eruptive flare we use the linear speed of the associated CME from the catalog. The SOHO/LASCO CME catalog does not list an entry for the M1.2/1N flare on 2011 October 1. Therefore we use the speed determined with STEREO-B in \cite{Temmer2017}.  
The GOES class, and therefore the peak GOES 1-8 \AA\ soft X-ray flux ($F_{\mathrm{SXR}}$), was taken from the flare reports from NGDC\footnote{\url{ftp://ftp.ngdc.noaa.gov/STP/space-weather/solar-data/solar-features/solar-flares/x-rays/goes/xrs/}}. $F_{\mathrm{SXR}}$ is used as an indicator of the flare energy.

\section{Methods}
\label{sec:methods}

\startlongtable
\begin{deluxetable*}{ccccccccc}
\tabletypesize{\footnotesize}
\renewcommand\arraystretch{0.89}
\renewcommand\tabcolsep{3.5pt}
\tablecaption{List of events. For each flare we  give the date and start/peak/end time, its heliographic position, the H$\alpha$ importance class (all from the KSO H$\alpha$ flare reports), and the GOES SXR flare class. The last column indicates whether the flare was associated with a CME or not (y/n).  \label{tab:events}}
\tablehead{\colhead{Event \#} & \colhead{Date} & \colhead{start time} & \colhead{peak time} & \colhead{end time} & \colhead{Heliographic} & \colhead{H$\alpha$\  importance} & \colhead{GOES SXR } & \colhead{CME} \\ 
 \colhead{} & \colhead{} & \colhead{(UT)} & \colhead{(UT)} & \colhead{(UT)} & \colhead{position} & \colhead{class} & \colhead{class} & \colhead{} } 
\startdata
1 & 2000 Jun 01 & 07:30 & 07:32 & 08:18 & S13 E24 & 2N & C 8.2 & n \\
2 & 2000 Jun 06 & 11:26 & 13:35\tablenotemark{b} & 13:52\tablenotemark{c} & N20 E12 & 3N & X 1.1 & n \\
3 & 2000 Jul 19 & 06:37\tablenotemark{a} & 07:23 & 09:01 & S15 E07 & 3N & M 6.4 & n \\
4 & 2000 Sep 12 & 11:22 & 12:00 & 14:58 & S19 W08 & 2F & M 1.0 & y \\
5 & 2001 Aug 25 & 16:23 & 16:32 & 17:25\tablenotemark{b} & S21 E38 & 3N & X 5.3 & y \\
6 & 2003 Oct 26 & 06:46\tablenotemark{a} & 06:46\tablenotemark{a} & 09:17 & S14 E41 & 3B & X 1.2 & y \\
7 & 2003 Oct 28 & 10:32\tablenotemark{a} & 11:23\tablenotemark{c} & 14:20 & S16 E07 & 4B & X17.2 & y \\
8 & 2003 Oct 29 & 20:37 & 20:42 & 22:53 & S15 W02 & 2B & X10.0 & y \\
9 & 2003 Nov 18 & 07:30\tablenotemark{a} & 07:50 & 11:04 & S02 E37 & 3N & M 3.2 & y \\
10 & 2003 Nov 20 & 07:35 & 07:42 & 08:43 & N01 W08 & 3B & M 9.6 & y \\
11 & 2004 Jul 16 & 13:50 & 13:57 & 14:31\tablenotemark{b} & S09 E29 & 4B & X 3.6 & n \\
12 & 2004 Jul 17 & 07:54 & 08:05 & 08:53 & S11 E22 & 2F & X 1.0 & n \\
13 & 2004 Jul 20 & 12:26 & 12:31 & 13:30 & N10 E32 & 3B & M 8.6 & y \\
14 & 2005 Jan 15 & 11:46 & 11:51 & 12:00 & N13 E01 & 2F & M 1.2 & n \\
15 & 2005 Jan 17 & 07:16\tablenotemark{a} & 09:51 & 11:57 & N14 W24 & 3B & X 3.8 & y \\
16 & 2005 May 12 & 07:28 & 07:34 & 08:57 & N12 E28 & 2B & M 1.6 & n \\
17 & 2005 Sep 12 & 08:42 & 08:49 & 11:05 & S13 E25 & 3N & M 6.1 & n \\
18 & 2005 Sep 15 & 08:34 & 08:40 & 10:10 & S11 W15 & 2N & X 1.1 & n \\
19 & 2006 Jul 06 & 08:16 & 08:42 & 10:24 & S10 W30 & 3N & M 2.5 & y \\
20 & 2011 Mar 07 & 13:48 & 14:31 & 14:50 & N10 E18 & 2F & M 1.9 & y \\
21 & 2011 Apr 22 & 11:09 & 11:33 & 12:02 & S16 E34 & 2N & C 7.7 & n \\
22 & 2011 Jun 02 & 07:25 & 07:47 & 08:11\tablenotemark{c} & S19 E20 & 2N & C 3.7 & y \\
23 & 2011 Sep 28 & 12:29 & 12:34 & 12:55 & N15 W01 & 1N & C 9.3 & n \\
24 & 2011 Oct 01 & 09:23 & 10:00 & 10:38 & N08 W03 & 1N & M 1.2 & y \\
25 & 2011 Nov 09 & 13:06 & 13:27 & 14:15 & N22 E36 & 2N & M 1.1 & y \\
26 & 2012 Mar 06 & 12:23 & 12:40 & 13:26 & N17 E35 & 2N & M 2.1 & n \\
27 & 2012 Mar 15 & 07:25 & 07:45 & 08:45 & N14 E00 & 2F & M 1.8 & y \\
28 & 2012 Apr 27 & 08:11 & 08:21 & 08:42 & N12 W30 & 1N & M 1.0 & n \\
29 & 2012 Jul 10 & 06:10 & 06:23 & 07:34 & S16 E30 & 1F & M 2.1 & n \\
30 & 2013 Apr 11 & 06:56 & 07:08 & 09:15 & N08 E14 & 3B & M 6.5 & y \\
31 & 2013 Jul 09 & 13:27 & 13:32 & 13:48 & S10 W21 & SN & C 2.3 & n \\
32 & 2013 Aug 02 & 11:10 & 11:11 & 11:24 & S15 W10 & SF & B 9.7 & n \\
33 & 2013 Aug 11 & 12:29 & 12:31 & 12:42 & S21 E31 & SF & B 7.1 & n \\
34 & 2013 Sep 23 & 07:10 & 07:11 & 07:24 & N10 E35 & SF & B 6.0 & n \\
35 & 2013 Oct 16 & 09:12 & 09:20 & 09:44 & S09 W42 & SF & C 1.9 & n \\
36 & 2013 Oct 20 & 08:36 & 08:41 & 09:08 & N22 W32 & 1N & C 2.9 & y \\
37 & 2013 Nov 29 & 09:55 & 10:10 & 10:14 & S06 W23 & 1F & C 1.5 & n \\
38 & 2013 Dec 14 & 11:06 & 11:19 & 11:58 & S14 W14 & 1F & C 2.3 & n \\
39 & 2013 Dec 28 & 12:42 & 12:44 & 13:05 & S17 E10 & 1F & C 3.0 & n \\
40 & 2014 Feb 14 & 10:38 & 10:40 & 11:04 & S11 W29 & 1N & C 7.2 & n \\
41 & 2014 Mar 21 & 10:18 & 10:35 & 11:01 & N17 E39 & 1F & C 2.7 & y \\
42 & 2014 May 02 & 09:17 & 09:23 & 10:19 & S19 W16 & 1N & C 4.4 & n \\
43 & 2014 May 10 & 06:51 & 07:01 & 08:02 & N03 E27 & 2N & C 8.7 & n \\
44 & 2014 May 12 & 06:25 & 06:38 & 07:07 & N04 W02 & 1F & C 2.3 & n \\
45 & 2014 Jun 21 & 13:36 & 13:54 & 14:03\tablenotemark{b} & S11 E04 & SF & B 4.7 & n \\
46 & 2014 Jun 26 & 07:12 & 07:31 & 07:46 & N10 E30 & SN & B 3.1 & n \\
47 & 2014 Aug 10 & 10:05 & 10:07 & 10:14 & S21 W12 & SF & B 8.9 & n \\
48 & 2014 Oct 22 & 14:02 & 14:06 & 14:55\tablenotemark{b} & S14 E15 & 3B & X 1.6 & n \\
49 & 2014 Oct 26 & 10:03\tablenotemark{a} & 10:51\tablenotemark{a} & 10:51\tablenotemark{b} & S14 W34 & 2N & X 2.0 & n \\
50 & 2014 Nov 02 & 13:07 & 13:11 & 13:19 & S04 E29 & SF & B 7.6 & n \\
51 & 2015 Jun 25 & 08:02 & 08:14 & 12:00 & N11 W41 & 3B & M 7.9 & y \\
\enddata
~\\
\tablenotetext{a}{Start/peak occurred before time listed}
\tablenotetext{b}{Event ended after time listed}
\tablenotetext{c}{Time is uncertain}
\end{deluxetable*}

This Section is divided into two parts. In Section \ref{subsec:pre-processing} we describe the pre-processing to obtain a homogeneous data set, which can be analyzed by our flare  detection algorithm. In Section \ref{subsec:analysis} we explain how we derived the magnetic reconnection quantities for the flares under study, illustrated on two sample events. 
\subsection{Pre-processing of H$\alpha$\ filtergrams}
\label{subsec:pre-processing}
The use of H$\alpha$\ filtergrams from different telescopes and CCDs makes a normalization of the data necessary. We apply a zero-mean and whitening transformation described in \cite{Poetzi2015}. We first normalize the intensity of each image using 
\begin{equation}
 f_{i,n}=\frac{f_i-\mu}{\sigma},
\end{equation}
where  $f_{i,n}$ is the normalized intensity and $f_{i}$ the original intensity of the i-th pixel of the image, $\mu$ the mean value of the intensity of all pixels on the solar disk defined as 
\begin{equation}
 \mu=\frac{1}{N}\sum_{i}^{N}f_{i},
\end{equation}
and $\sigma$ the corresponding standard deviation, defined as
\begin{equation}
 \sigma=\sqrt{\frac{1}{N-1}\sum_{i}^{N}\left(f_i-\mu\right)^2},
\end{equation}
with $N$ the number of pixels on the solar disk. This means, the normalized intensity is given in units of the standard deviation of the solar disk pixels. 

The area covered by the flare ribbons is small compared to the area of the solar disk, and has thus no significant impact on the mean value $\mu$. After normalization, a median filter is applied to eliminate large scale intensity variations such as the center to limb variation and the effect of terrestrial clouds. This results in the normalized intensity values of the quiet sun to be close to zero. All features darker than the quiet Sun show negative values while brighter features have  positive values of the normalized intensity. 

All images are rotated to solar north and differentially rotated to the time of the first image of the H$\alpha$\ image sequence used to analyze the individual flare events. For each event a subregion containing the flare ribbons is selected. We co-align the H$\alpha$\ images with the first image of the time series using cross correlation techniques to compute the offsets, and shift the images accordingly. The SOHO MDI and SDO HMI LOS magnetograms are co-registered to the plate scale of the H$\alpha$\ images and then co-aligned with the first 
H$\alpha$\ filtergram of each event-based image sequence, using the corresponding MDI or HMI continuum images, respectively. 

\subsection{Analysis of reconnection rates and fluxes}
\label{subsec:analysis}

In each H$\alpha$\ filtergram of the observation series we detect those pixel, which belong to the flare ribbons. We define all pixels as flare pixels whose normalized intensity is $\ge$5.5, i.e. the pixel intensity lies 5.5$\sigma$ above the mean value of the
solar disk pixel intensities. The normalization allows us to use the same threshold for all events. The threshold is found by visual inspection of the flare pixel tracking method, based on different thresholds, in comparison with the flare ribbon area observed in the H$\alpha$\ filtergrams. Within a range of [4.5, 6.5] in normalized intensity, qualitatively similar detection results are obtained. Therefore we use a value of 5.5 of the normalized intensity as threshold for flare pixel detection, and values of 4.5 and 6.5 to assign a lower and upper uncertainty bound, respectively. 

\begin{figure}
 \begin{center}
  \includegraphics[width=0.75\textwidth]{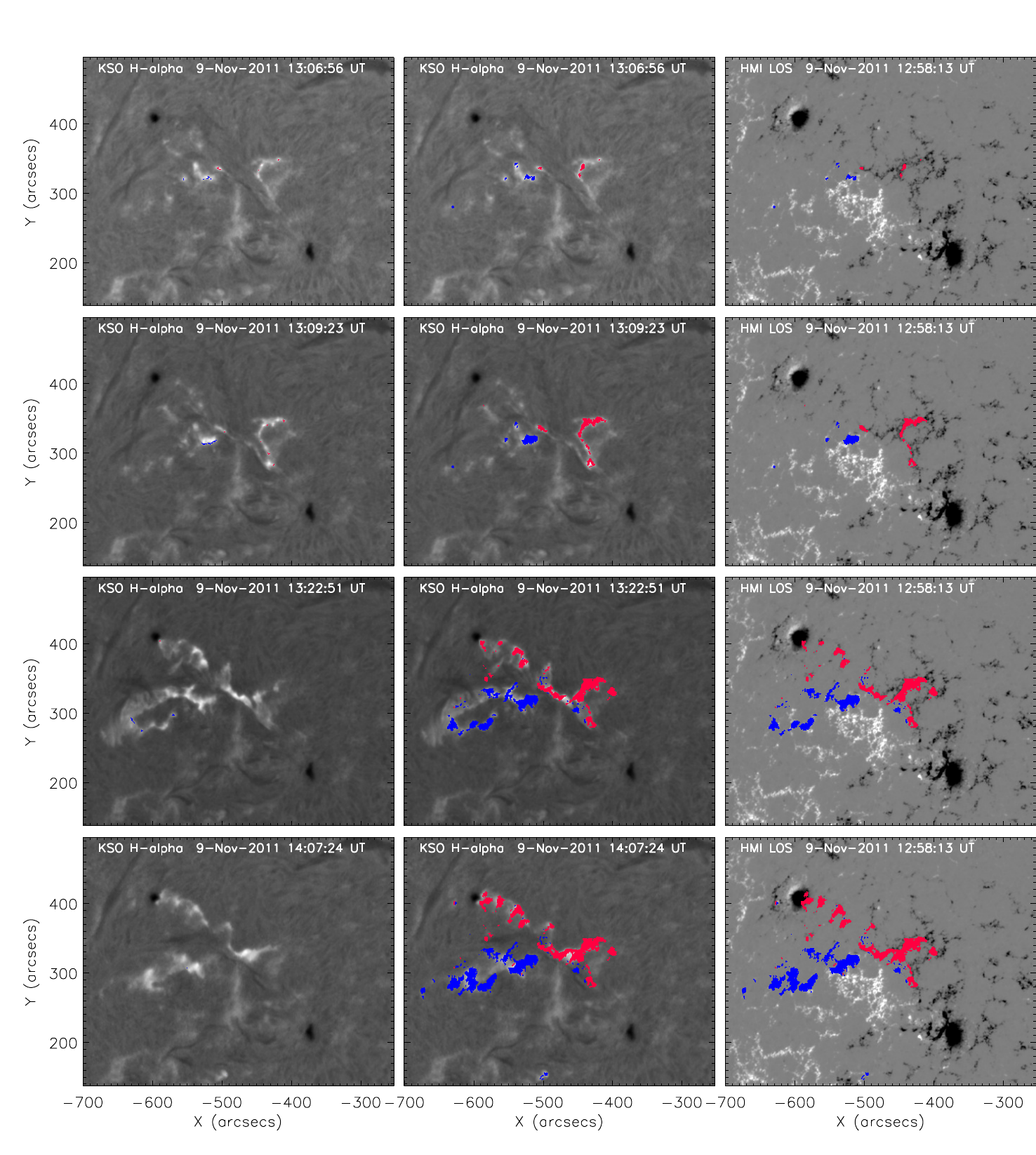}
  \caption{Overview of the evolution of the eruptive M1.1 flare on 2011 November 9. The left column shows the newly brightened flare pixels on top of the KSO H$\alpha$ image. The middle column shows the cumulated flare pixels up to the time of the image shown. The right column shows the cumulated flare flare pixels on top of the HMI LOS magnetogram. The HMI magnetogram is scaled to $\pm$500 G. Red areas indicate negative magnetic polarity, blue areas positive magnetic polarity of the flare ribbons. A movie of the time evolution is available online in the supplementary materials.}
  \label{fig:sequence_er}
 \end{center}
\end{figure}

\begin{figure}
  \begin{center}
  \includegraphics[width=0.75\textwidth]{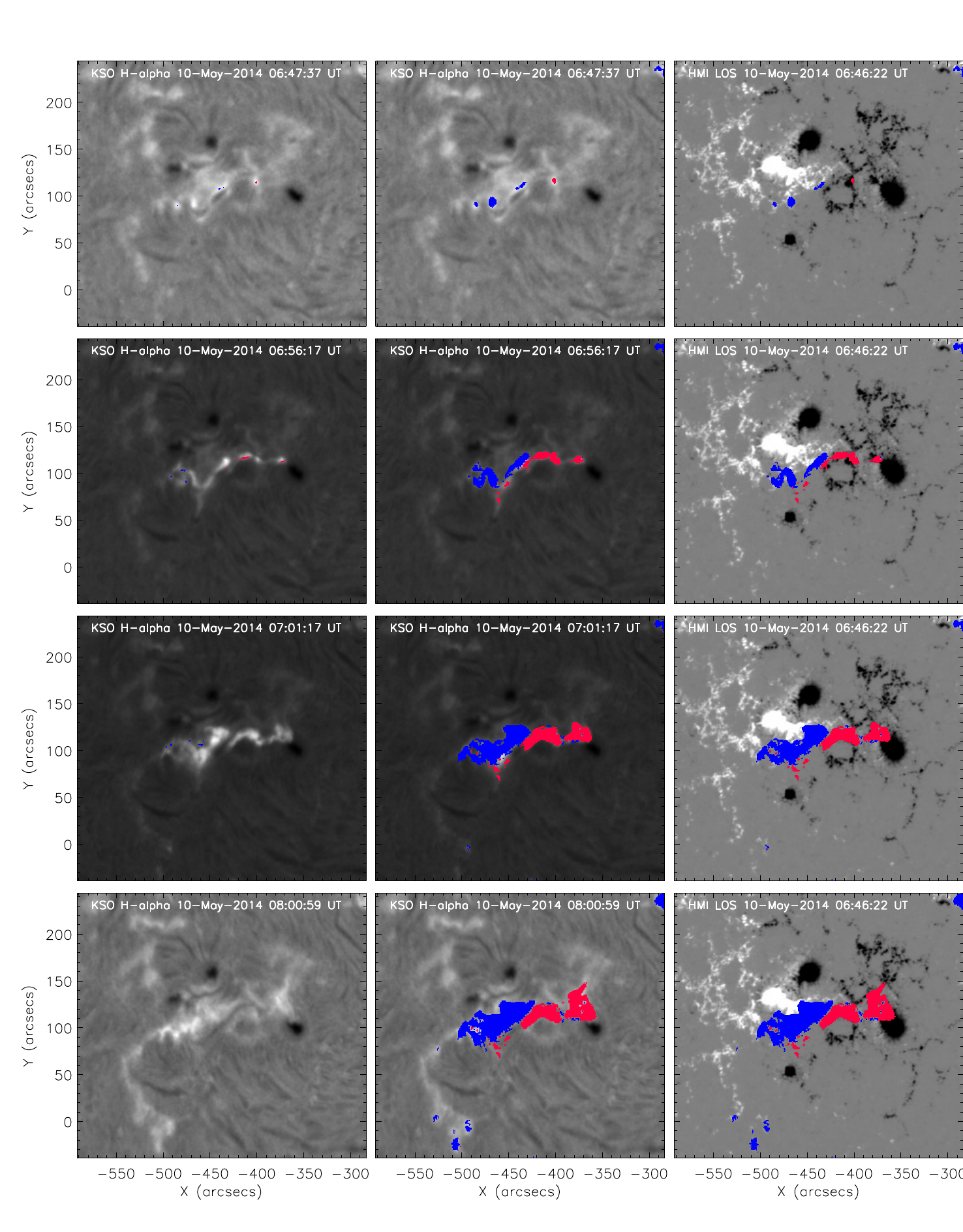}
  \caption{Overview of the evolution of the confined C8.7 flare on 2014 May 10. The left column shows the newly brightened flare pixels on top of the KSO H$\alpha$ image. The middle column shows the cumulated flare pixels up to the time of the image shown. The right column shows the cumulated flare flare pixels on top of the HMI LOS magnetogram. The HMI magnetogram is scaled to $\pm$500 G. Red areas indicate negative magnetic polarity, blue areas positive magnetic polarity of the flare ribbons. A movie of the time evolution is available online in the supplementary materials.}
  \label{fig:sequence_con}
 \end{center}
\end{figure}

\begin{figure}
 \begin{center}
  \includegraphics[width=0.75\textwidth]{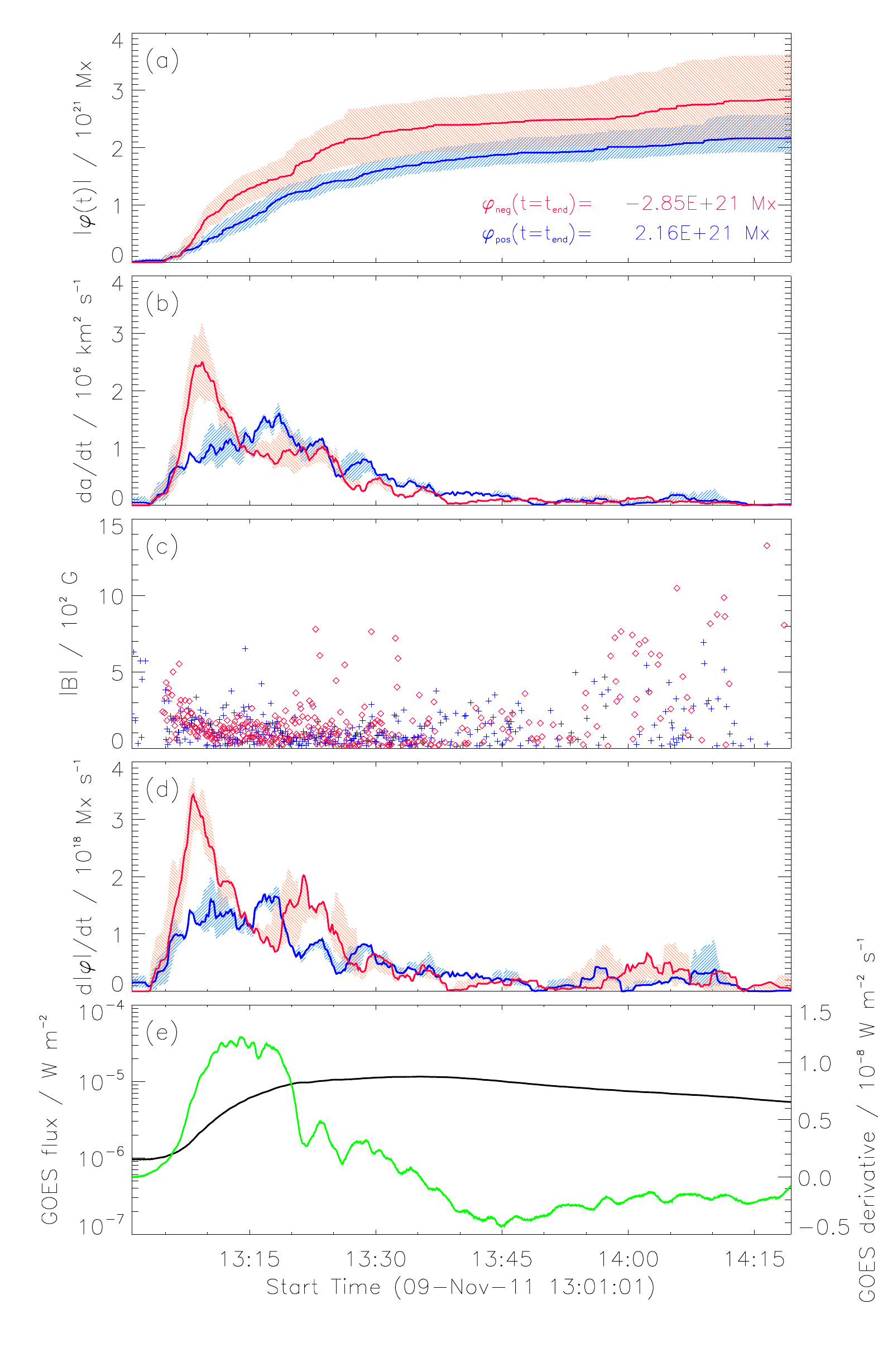}
  \caption{Time evolution of the calculated magnetic reconnection parameters for the eruptive M1.1 flare on 2011 November 9. (a) Flare reconnection flux $\varphi\left(t\right)$, (b) rate of change in flare area, (c) mean magnetic field in the newly brightened flare pixels, (d) reconnection rate $\dot{\varphi}\left(t\right)$. The thick lines indicate the derived values, the shaded regions denote the uncertainty ranges. The red curves are derived from negative polarity regions, while blue curves are derived from positive polarity regions. (e) GOES 1-8 \AA\ soft X-ray flux (black) and its time derivative (green).}
  \label{fig:time_evolution_er}
 \end{center}
\end{figure}

\begin{figure}
 \begin{center}
  \includegraphics[width=0.75\textwidth]{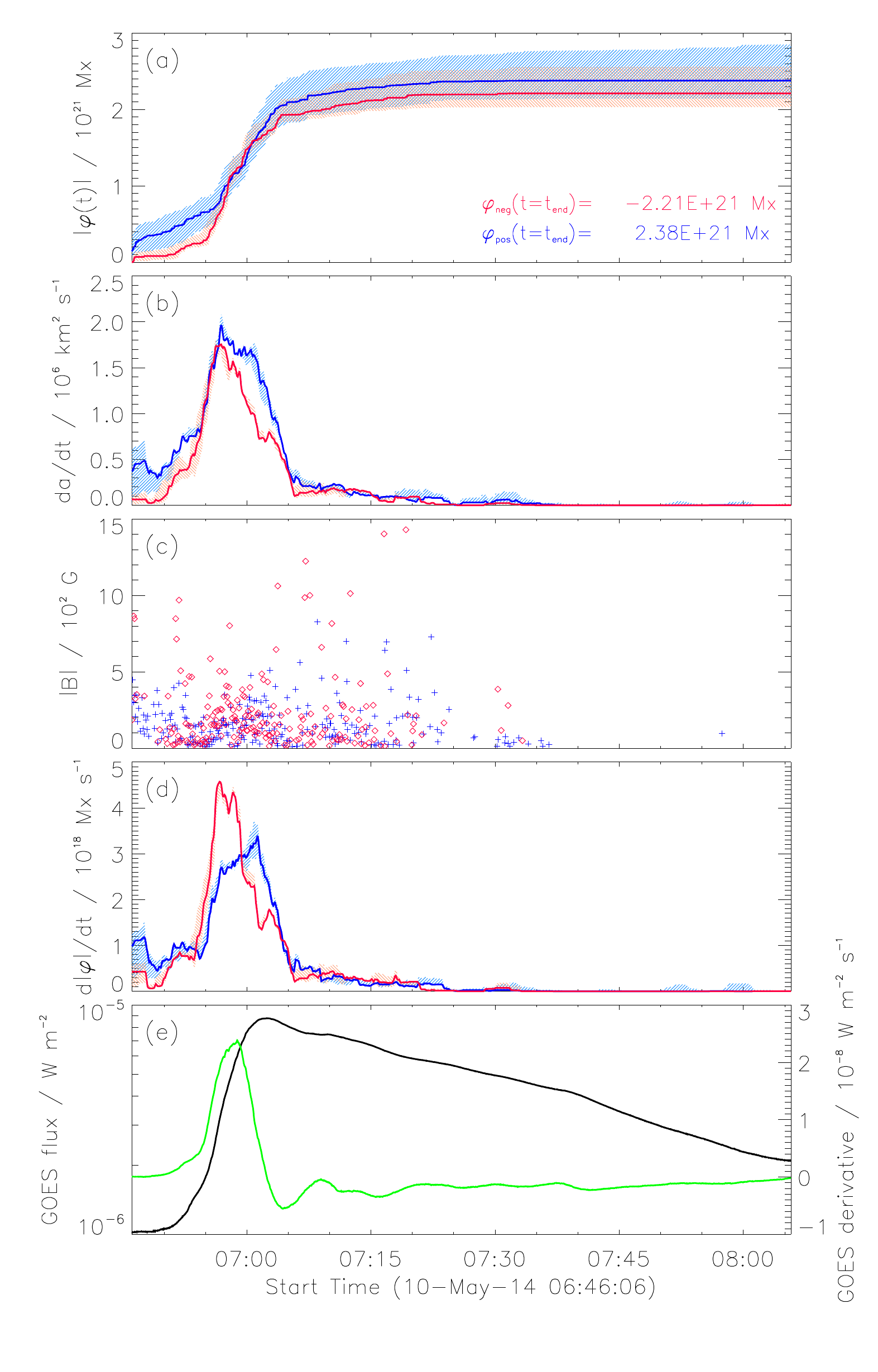}
  \caption{Time evolution of the calculated reconnection parameters for the confined C8.7 flare on 2014 May 10. (a) Flare reconnected flux $\varphi\left(t\right)$, (b) rate of change in flare area, (c) mean magnetic field in the newly brightened flare pixels, (d) reconnection rate $\dot{\varphi}\left(t\right)$. The thick lines indicate the derived values, the shaded regions denote the uncertainty ranges. The red curves are derived from negative polarity regions, while blue curves are derived from positive polarity regions. (e) GOES 1-8 \AA\ soft X-ray flux (black) and its time derivative (green).}
  \label{fig:time_evolution_con}
 \end{center}
\end{figure}

Figures \ref{fig:sequence_er} and \ref{fig:sequence_con} illustrate the detection of flare pixels during the eruptive M1.1/2N flare on 2011 November 9 (event \#25) and the confined C8.7/2N flare on 2014 May 10 (event \#43), respectively. The left column shows the newly detected flare pixels between the image shown and the previous one (recorded 6 seconds earlier), while the middle column shows the accumulated ribbon area up to the time of the image recorded. The right column shows the accumulated flare area on top of a pre-flare HMI LOS magnetogram. Areas colored in red and blue correspond to flare ribbons populating regions of negative and positive polarity, respectively. With time the accumulated ribbon area grows, with the largest growth happening during the impulsive phase of the flare and slowing down in the later phases of the flare. This is displayed in Figures \ref{fig:time_evolution_er}b and \ref{fig:time_evolution_con}b, showing the growth rates of the ribbon area during the two sample vents.

For each time step the magnetic flux of the newly brightened flare pixels is calculated. A pixel is considered a newly brightened flare pixel belonging to a certain magnetic polarity if a) the normalized intensity of the pixel exceeds a threshold of 5.5, b) this intensity threshold has not been exceeded in any of the previous H$\alpha$ filtergrams, and c) the absolute value of the LOS magnetic flux density is above the noise level. The noise level used is 20 G in case of the MDI magnetograms \citep{Scherrer1995}. For HMI 720s LOS magnetograms the photon noise is 3 G at the disk center  \citep{Couvidat2016}. A noise level of 10 G is used to account for the increase of noise towards the limb.

The magnetic flux in the newly brightened flare pixels is then used to evaluate the reconnection flux. First we calculate the difference in the reconnection flux between the time step $t_{k}$ and $t_{k-1}$ as
\begin{equation}
 \Delta\varphi\left(t_k\right)=\sum_{i}a_{i}\left(t_{k}\right)B_{n}\left(a_{i}\right)
 \label{eq:phi}
\end{equation}
where $a_{i}\left(t_{k}\right)$ is the area of the flare pixels that newly brightened between time step $t_{k-1}$ and $t_{k}$. 
$B_{n}\left(a_{i}\right)$ is the component of the magnetic field normal to the surface in the newly brightened area. $B_{n}$ is estimated by multiplying the LOS magnetic flux density with $1/\cos(u)$, where $u$ is the angular distance of the centroid of all flare pixels to the center of the solar disk. Likewise, the calculated flare area is corrected for projection effects by multiplying with $1/\cos(u)$. The calculation of $\Delta\varphi\left(t_k\right)$ is done separately for the positive and negative polarity domains. From this sequence, we obtain the cumulated reconnection flux up to time $t_{k}$ as
\begin{equation}
 \varphi\left(t_{k}\right)=\sum_{j\le k}\Delta\varphi\left(t_{j}\right).
 \label{eq:sum}
\end{equation}
In Eq.~\ref{eq:sum}, the reconnection flux $\varphi\left(t_{k}\right)$ is derived as the sum of all fluxes in the flare areas that brightened up until  time $t_{k}$. The time series is smoothed with a three minute time window. The reconnection rate, i.e. the change rate of the magnetic reconnection flux at each time step, is then calculated as \citep[cf.][]{Veronig2015}
\begin{equation}
 \dot{\varphi}\left(t_{k}\right)=\frac{\Delta\varphi\left(t_{k}\right)}{\left(t_{k}-t_{k-1}\right)}.
\end{equation}

Figures \ref{fig:time_evolution_er} and \ref{fig:time_evolution_con} show the time evolution of all the  parameters calculated, along with the GOES 1-8 \AA\ flux and its time derivative for the sample events \#25 and \#43, respectively. 
Figures \ref{fig:time_evolution_er}a and \ref{fig:time_evolution_con}a show the reconection flux $\varphi\left(t\right)$, separately for positive and negative magnetic flux areas (thick lines) with the 
shaded regions indicating the uncertainty range obtained
by the flare detection thresholds of [4.5, 6.5]. The reconnection flux increases with time, while showing small differences in positive and negative flux. Figures \ref{fig:time_evolution_er}b and \ref{fig:time_evolution_con}b show the corresponding growth rate of the flare ribbon areas. Note that the ribbon areas covering opposite-polarity regions do not necessarily grow homogeneously, as is e.g. the case in event \#25. Figures \ref{fig:time_evolution_er}c and \ref{fig:time_evolution_con}c show the mean magnetic flux density in the newly brightened flare areas. Figures \ref{fig:time_evolution_er}d and \ref{fig:time_evolution_con}d show the evolution of the reconnection rate $\dot{\varphi}\left(t\right)$. 

We define the total flare reconnection flux $\varphi_{\mathrm{FL}}$ as the mean of the absolute values of the reconnection fluxes in both polarity regions at the end of the time series, i.e.
\begin{equation}
 \varphi_{\mathrm{FL}}=\frac{\abs{\varphi_{+}}+\abs{\varphi_{-}}}{2},
 \label{eq:phitot}
\end{equation}
with $\varphi_{+}$ and $\varphi_{-}$ the cumulated magnetic flux in the positive and negative polarity regions, respectively. Accordingly, the peak reconnection rate is calculated as
\begin{equation}
 \dot{\varphi}_{\mathrm{FL}}=\frac{\abs{\dot{\varphi}_{+}}+\abs{\dot{\varphi}_{-}}}{2},
\end{equation}
with $\dot{\varphi}_{+}$ and $\dot{\varphi}_{-}$ the peaks of the reconnection rate within positive and negative magnetic polarity, respectively. The uncertainties of these quantities due to a particular choice of threshold were estimated by performing the analysis described above based on intensity thresholds [4.5, 6.5]. We find, that these uncertainties are about 20--30\%. 
The effects of these uncertainties in the alignment of the images was estimated by randomly misaligning each image by up to 2 pixels in both, either positive or negative, x- and y-direction and running the calculations for the misaligned images with a threshold of 5.5. We find that the uncertainties in the reconnection flux due to inaccurate alignment is about 10\%. In total we therefore estimate the maximum uncertainties in the derived reconnection fluxes to be 30--40\%.

In order to compare the flare reconnection flux to the total magnetic flux contained in the source active region (AR), we selected a sub-region containing the whole AR. This box was chosen as small as possible around the AR, to minimize the contribution of magnetic flux from areas outside the source AR. 
We used the same pre-flare LOS magnetogram that was used for the calculation of the flare reconnections fluxes, and calculated the flux in the sub-region selected to provide an estimate of the total magnetic flux contained in the flare-hosting AR.
Three of our flares did not originate from ARs (events \#1, \#4 and \#36). In these cases we used the same sub-region which was used in the analysis of the flare reconnection rates. The flux within each polarity was calculated separately, and analogously to Eq.~\ref{eq:phitot} we defined a total active region flux $\varphi_{\mathrm{AR}}$ as the mean of the absolute values of the magnetic flux of both polarity regions. 
The errors in the calculation of the magnetic flux in the AR that causes the flare are estimated to be about 5\%.  They mostly arises due to the selection of the box around the AR from where the flux is derived, as it may either include also parts of quiet Sun fluxes around the AR or may miss some flux from the periphery of the AR. In both cases, these contributions are small as they cover only regions with small magnetic flux densities, whereas most of the AR flux stems from the umbra and penumbra of the main sunspots.

The partition of the AR flux involved in the flare reconnection process was estimated as
\begin{equation}
 r=\frac{\varphi_{\mathrm{FL}}}{\varphi_{\mathrm{AR}}}.
\end{equation}
Combining the relative error bounds of 30\% for $\varphi_{\mathrm{FL}}$ and 5\% for $\varphi_{\mathrm{AR}}$, we find a maximum estimate on the relative error on $r$  of about 35\%. 

\section{Results}
\label{sec:results}

In Table \ref{tab:results}, we list the reconnection parameters derived for all the flare events under study. The reconnection fluxes $\varphi_{\mathrm{FL}}$ derived for our sample range over more than two orders of magnitude, 
with a minimum of $\varphi_{\mathrm{FL}}=1.7\cdot 10^{20}\, \mathrm{Mx}$ for event \#50 and a maximum of ${\varphi_{\mathrm{FL}}=2.5\cdot 10^{22}\, \mathrm{Mx}}$ for event \#7, while $F_{\mathrm{SXR}}$ covers a range of 4 orders of magnitude from B3 to X17 (cf. Table \ref{tab:results} and Figure \ref{fig:phi_goes_sep}). The peak reconnection rates $\dot{\varphi}_{\mathrm{FL}}$ are also distributed over two
orders of magnitude ranging from ${\dot{\varphi}_{\mathrm{FL}}=5.2\cdot 10^{17}\, \mathrm{Mx\,s^{-1}}}$ in case of event \#50 to ${\dot{\varphi}_{\mathrm{FL}}=3.4\cdot 10^{19}\, \mathrm{Mx\,s^{-1}}}$ in case of event \#8.

\begin{figure}
 \includegraphics[width=\textwidth]{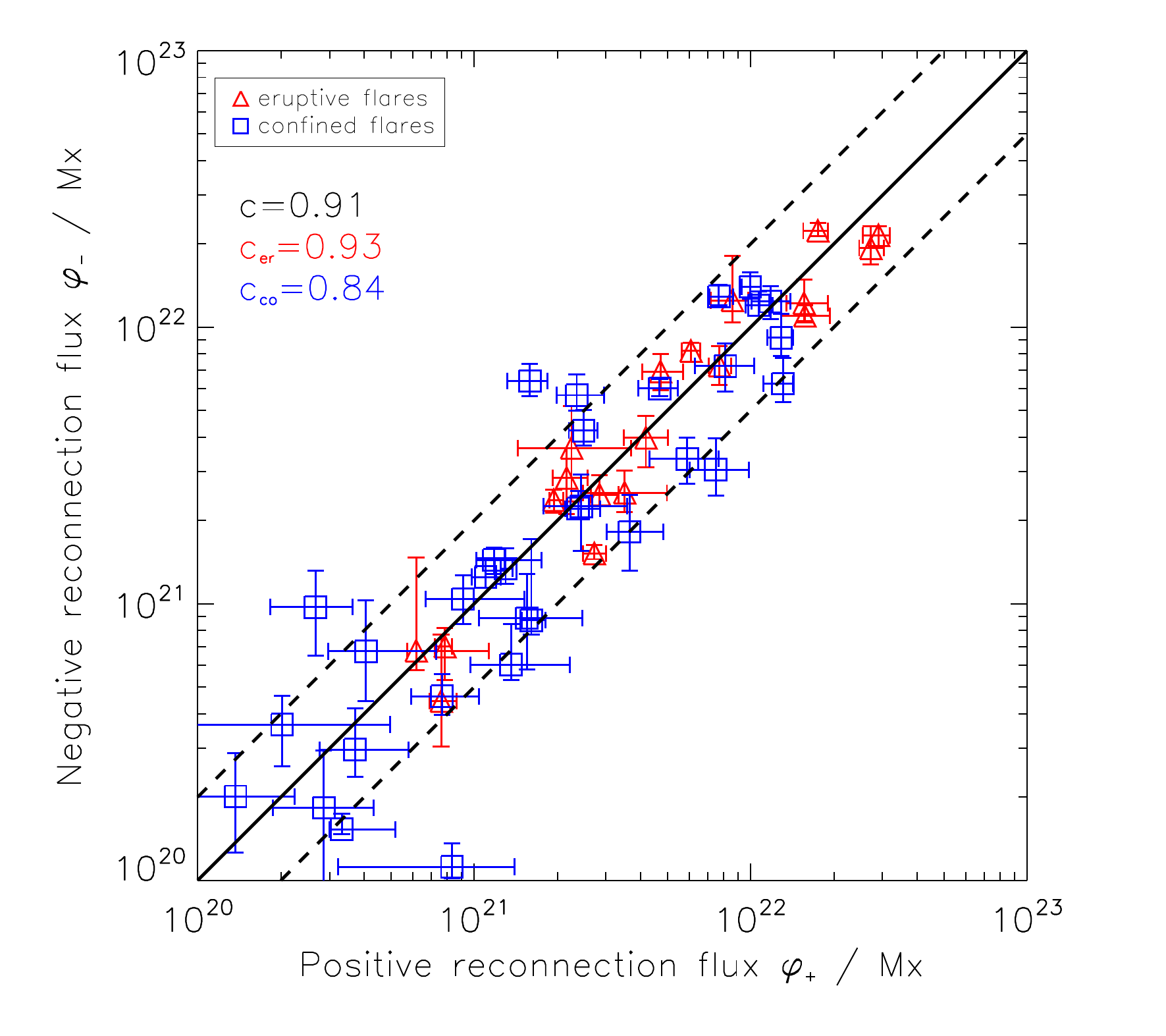}
 \caption{Negative ($\varphi_{-}$) against positive ($\varphi_{+}$) flare reconnection flux. Blue squares indicate confined, red triangles eruptive events. The black line indicates a 1:1 relation between the reconnected flux in opposite magnetic polarities. The dashed lines indicate ratios of 0.5 and 2 between the positive and negative fluxes.}
 \label{fig:phi_phi}
\end{figure}

Figure \ref{fig:phi_phi} shows the negative versus positive magnetic flux reconnected during each of the events analyzed.  In most events, the reconnected magnetic flux in the two polarities reside near the black line indicating $r_{\mathrm{pn}}=1$, indicating perfect flux balance in the opposite polarities. The dashed lines represent the lines where $r_{\mathrm{pn}}$ is either 0.5 or 2. For the majority of events (82\%), we find ratios of $r_{\mathrm{pn}}=\abs{\varphi_{+}/\varphi_{-}}$ within a range of 0.5 and 2. Considering the uncertainties in the measurements, this is a good flux balance.  Six events show a ratio $r_{\mathrm{pn}}$ larger than 2, while three events exhibit $r_{\mathrm{pn}}$ smaller than 0.5.  

Figure \ref{fig:phi_goes_sep} shows the total reconnection flux, $\varphi_{\mathrm{FL}}$, as a function of the peak of the GOES 1-8 \AA\ SXR flux, $F_{\mathrm{SXR}}$, indicating an 
excellent correlation with a Pearson correlation ceofficient
of  $c=0.92$, derived in log-log space. This is true also for the subsets of confined and eruptive flares, with the corresponding fit parameters showing no significant differences within their given uncertainties. For the whole set of events we find values of $a=0.580$ and $d=24.21$ for the regression line $\log\left(y\right)=a\,\log\left(x\right)+d$ (see first column in Table \ref{tab:fitcc}).   
The coefficient of determination for our regression to the data (black line in Fig. \ref{fig:phi_goes_sep}) is given by $c^2 = 0.85$. The coefficient of determination provides us with a measure how well the regression line can account for the variation of the data. In the case of the relation $\varphi_{\mathrm{FL}}$ versus $F_{\mathrm{SXR}}$, 85\% of the total variation of the flare reconnection fluxes can be accounted for by the regression model, confirming a high applicability of the regression model. 

Figure \ref{fig:scatter_plots}a shows $\dot{\varphi}_{\mathrm{FL}}$ as a function of the peak of the time derivative of the SXR flux, $\dot{F}_{\mathrm{SXR}}$. These quantities also show a very high correlation with $c=0.90$. We find values for the fit parameters of $a=0.456$ and $d=22.03$ (see third column in Table \ref{tab:fitcc}), with no significant differences in the fit parameters when considering confined and eruptive flares separately. Figure \ref{fig:scatter_plots}b shows that $\dot{\varphi}_{\mathrm{FL}}$ is also strongly correlated with the SXR peak flux, $F_{\mathrm{SXR}}$ ($c=0.86$). There is some trend that for a given GOES flare class ($F_{\mathrm{SXR}}$), confined flares reveal higher peak reconnection rates $\dot{\varphi}_{\mathrm{FL}}$  than eruptive flares. 

Figures \ref{fig:ar_b_phi}a and \ref{fig:ar_b_phi}b show $\varphi_{\mathrm{FL}}$ as a function of the flare area $A_{\mathrm{FL}}$ and the mean magnetic flux density $\bar{\abs{B}}_{\mathrm{FL}}$ in the flare ribbons, respectively. 
We find very high correlation coefficients for both eruptive and confined flares ($c \approx 0.9$), but a difference in the regression lines. 
For a given reconnection flux $\varphi_{\mathrm{FL}}$, confined flares involve a higher mean magnetic flux density $\bar{\abs{B}}_{\mathrm{FL}}$ and a smaller area as compared to eruptive flares. The effect is most pronounced in the magnetic field underlying the flare ribbons, which is about a factor of 2 larger in case of confined flares than in eruptive events, quite constant over the whole distribution.
(Note that the flare area  $A_{\mathrm{FL}}$ and the mean magnetic flux density $\bar{\abs{B}}_{\mathrm{FL}}$ in the flare ribbons themselves are only very weakly correlated, $c = 0.34 \,(-0.26\, / +0.21)$).

\startlongtable
\begin{deluxetable*}{cccccccc}
\tabletypesize{\footnotesize}
\renewcommand\arraystretch{0.95}
\renewcommand\tabcolsep{3.5pt}
\tablecaption{Flare reconnection fluxes $\varphi$, reconnection rates $\dot{\varphi}$ and peak of GOES 1-8 \AA\ SXR flux derivate $\dot{F}_{\mathrm{SXR}}$. \label{tab:results}}
\tablehead{\colhead{Event \#} & \colhead{$\varphi_{+}$} & \colhead{$\varphi_{-}$} & \colhead{$\varphi_{\mathrm{FL}}$} & \colhead{$\dot{\varphi}_{+}$} & \colhead{$\dot{\varphi}_{-}$} & \colhead{$\dot{\varphi}_{\mathrm{FL}}$} & \colhead{$\dot{F}_{\mathrm{SXR}}$} \\ 
\colhead{} & \colhead{($10^{21}\, \mathrm{Mx}$)} & \colhead{($10^{21}\, \mathrm{Mx}$)} & \colhead{($10^{21}\, \mathrm{Mx}$)} & \colhead{($10^{19}\, \mathrm{Mx\,s}^{-1}$)} & \colhead{($10^{19}\, \mathrm{Mx\,s}^{-1}$)} & \colhead{($10^{19}\, \mathrm{Mx\,s}^{-1}$)} & \colhead{($10^{-8}\, \mathrm{W}\,\mathrm{m}^{2}\,\mathrm{s}^{-1}$)} } 
\startdata
1 &  1.32 &  -1.37 &  1.34 & 0.421 & -0.400 & 0.410 & 3.940 \\
2 &  7.68 &  -12.9 &  10.3 & 0.543 & -0.737 & 0.640 & 40.80 \\
3 &  11.8 &  -12.5 &  12.1 & 0.923 & -0.600 & 0.761 & 4.680 \\
4 &  2.27 &  -3.68 &  2.97 & 0.127 & -0.152 & 0.139 & 0.696 \\
5 &  15.8 &  -11.1 &  13.4 &  1.56 &  -1.51 &  1.53 & 138.0 \\
6 &  15.6 &  -12.2 &  13.9 & 0.625 & -0.922 & 0.773 & 2.910 \\
7 &  29.0 &  -21.5 &  25.2 &  3.94 &  -2.86 &  3.40 & 508.0 \\
8 &  27.2 &  -19.4 &  23.3 &  4.58 &  -2.30 &  3.44 & 302.0 \\
9 &  6.18 &  -8.24 &  7.21 & 0.307 & -0.377 & 0.342 & 4.290 \\
10 &  4.74 &  -6.95 &  5.84 & 0.641 &  -1.89 &  1.27 & 29.70 \\
11 &  13.1 &  -6.28 &  9.69 &  2.39 &  -1.50 &  1.94 & 209.0 \\
12 &  5.92 &  -3.35 &  4.63 &  1.71 & -0.980 &  1.34 & 66.00 \\
13 &  8.56 &  -12.4 &  10.5 &  2.27 &  -3.84 &  3.05 & 37.00 \\
14 &  8.10 &  -7.24 &  7.67 &  1.24 & -0.537 & 0.888 & 4.650 \\
15 &  17.5 &  -22.4 &  19.9 &  1.14 & -0.658 & 0.899 & 45.60 \\
16 &  2.51 &  -4.24 &  3.37 & 0.744 & -0.760 & 0.752 & 8.300 \\
17 &  12.9 &  -9.19 &  11.0 &  1.68 &  -1.73 &  1.70 & 20.10 \\
18 &  10.7 &  -12.0 &  11.3 &  3.34 &  -3.16 &  3.25 & 62.10 \\
19 &  4.28 &  -4.08 &  4.18 & 0.554 & -0.342 & 0.448 & 5.830 \\
20 &  3.50 &  -2.52 &  3.01 & 0.548 & -0.358 & 0.453 & 2.310 \\
21 &  1.59 &  -6.39 &  3.99 & 0.232 &  -1.01 & 0.621 & 1.550 \\
22 & 0.617 & -0.676 & 0.646 & 0.101 & -0.097 & 0.099 & 0.516 \\
23 &  1.18 &  -1.44 &  1.31 & 0.439 & -0.731 & 0.585 & 5.650 \\
24 &  1.94 &  -2.37 &  2.15 & 0.175 & -0.148 & 0.161 & 1.320 \\
25 &  2.16 &  -2.85 &  2.50 & 0.169 & -0.343 & 0.256 & 1.220 \\
26 &  4.69 &  -6.01 &  5.35 & 0.561 & -0.587 & 0.574 & 4.670 \\
27 &  2.72 &  -1.52 &  2.12 & 0.319 & -0.188 & 0.253 & 5.350 \\
28 &  1.10 &  -1.25 &  1.17 & 0.241 & -0.218 & 0.229 & 4.390 \\
29 &  2.35 &  -5.68 &  4.01 & 0.426 & -0.804 & 0.615 & 3.750 \\
30 &  2.84 &  -2.48 &  2.66 & 0.370 & -0.411 & 0.390 & 10.40 \\
31 & 0.765 & -0.463 & 0.614 & 0.188 & -0.176 & 0.182 & 1.240 \\
32 & 0.286 & -0.183 & 0.234 & 0.125 & -0.035 & 0.080 & 0.310 \\
33 & 0.267 & -0.973 & 0.620 & 0.095 & -0.206 & 0.151 & 0.161 \\
34 & 0.371 & -0.297 & 0.334 & 0.143 & -0.143 & 0.143 & 0.194 \\
35 &  1.36 & -0.602 & 0.981 & 0.424 & -0.207 & 0.315 & 0.446 \\
36 & 0.784 & -0.700 & 0.742 & 0.181 & -0.111 & 0.146 & 1.160 \\
37 & 0.332 & -0.153 & 0.242 & 0.144 & -0.035 & 0.089 & 0.429 \\
38 &  1.55 & -0.888 &  1.22 & 0.195 & -0.110 & 0.152 & 0.309 \\
39 &  1.61 & -0.872 &  1.24 & 0.573 & -0.320 & 0.446 & 1.460 \\
40 &  3.65 &  -1.82 &  2.73 & 0.894 & -0.562 & 0.728 & 4.350 \\
41 & 0.760 & -0.445 & 0.603 & 0.069 & -0.056 & 0.062 & 0.194 \\
42 &  2.44 &  -2.25 &  2.34 & 0.298 & -0.340 & 0.319 & 0.580 \\
43 &  2.38 &  -2.21 &  2.29 & 0.339 & -0.457 & 0.398 & 2.390 \\
44 & 0.912 &  -1.04 & 0.976 & 0.114 & -0.209 & 0.161 & 0.271 \\
45 & 0.406 & -0.673 & 0.539 & 0.119 & -0.136 & 0.127 & 0.097 \\
46 & 0.831 & -0.112 & 0.471 & 0.146 & -0.028 & 0.087 & 0.068 \\
47 & 0.202 & -0.366 & 0.284 & 0.047 & -0.110 & 0.078 & 0.285 \\
48 &  9.97 &  -14.0 &  12.0 &  1.74 &  -2.38 &  2.06 & 24.90 \\
49 &  7.48 &  -3.05 &  5.26 & 0.844 & -0.419 & 0.631 & 2.350 \\
50 & 0.137 & -0.201 & 0.169 & 0.058 & -0.046 & 0.052 & 0.158 \\
51 &  7.71 &  -7.26 &  7.48 & 0.580 & -0.790 & 0.685 & 39.90 \\
\enddata
~\\
\end{deluxetable*}

\startlongtable
\begin{deluxetable*}{lcccccccccc}
\tabletypesize{\footnotesize}
\tablecaption{Fit parameters and correlation coefficients corresponding to the scatter plots in Figures \ref{fig:phi_goes_sep}--\ref{fig:ar_b_goes}. The fit parameters and correlation coefficients are calculated in log-log space using $ \log \left(y\right)=a \log\left(x\right)+d$, for all events as well as separately for the subsets of confined and eruptive events.  \label{tab:fitcc}}
\tablehead{\colhead{x} & \multicolumn{2}{c}{$F_{\mathrm{SXR}}$} & \colhead{} & \colhead{$\dot{F}_{\mathrm{SXR}}$} & \colhead{} & \multicolumn{2}{c}{$A_{\mathrm{FL}}$} & \colhead{} & \multicolumn{2}{c}{$\bar{\abs{B}}_{\mathrm{FL}}$} \\ \cline{2-3} \cline{5-5} \cline{7-8} \cline{10-11}
\colhead{y} & \colhead{$\varphi_{\mathrm{FL}}$} & \colhead{$\dot{\varphi}_{\mathrm{FL}}$} & \colhead{} & \colhead{$\dot{\varphi}_{\mathrm{FL}}$} & \colhead{} & \colhead{$\varphi_{\mathrm{FL}}$} & \colhead{$F_{\mathrm{SXR}}$} & \colhead{} & \colhead{$\varphi_{\mathrm{FL}}$} & \colhead{$F_{\mathrm{SXR}}$} } 
\startdata
$a$ & 0.580$\pm$0.034 & 0.457$\pm$0.038 &   & 0.457$\pm$0.032 &   & 1.207$\pm$0.071 & 1.845$\pm$0.137 &   & 1.709$\pm$0.245 & 2.284$\pm$0.444 \\
$a_{\mathrm{con}}$ & 0.602$\pm$0.052 & 0.484$\pm$0.045 &   & 0.467$\pm$0.041 &   & 1.259$\pm$0.082 & 1.778$\pm$0.170 &   & 1.959$\pm$0.305 & 2.404$\pm$0.553 \\
$a_{\mathrm{er}}$ & 0.579$\pm$0.050 & 0.577$\pm$0.060 &   & 0.501$\pm$0.048 &   & 1.705$\pm$0.144 & 2.525$\pm$0.365 &   & 1.833$\pm$0.170 & 2.925$\pm$0.312 \\ \hline
$d$ & 24.21$\pm$0.17 & 20.82$\pm$0.19 &   & 22.03$\pm$0.24 &   & 10.23$\pm$0.66 & -21.92$\pm$1.28 &   & 17.23$\pm$0.60 & -10.41$\pm$1.09 \\
$d_{\mathrm{con}}$ & 24.34$\pm$0.27 & 21.05$\pm$0.23 &   & 22.16$\pm$0.32 &   & 9.82$\pm$0.74 & -21.28$\pm$1.55 &   & 16.42$\pm$0.76 & -11.07$\pm$1.38 \\
$d_{\mathrm{er}}$ & 24.17$\pm$0.22 & 21.20$\pm$0.27 &   & 22.26$\pm$0.34 &   & 5.36$\pm$1.37 & -28.49$\pm$3.49 &   & 17.27$\pm$0.41 & -11.35$\pm$0.75 \\ \hline
$c$ & 0.92(-0.05/0.03) & 0.86(-0.09/0.06) &   & 0.90(-0.07/0.04) &   & 0.93(-0.05/0.03) & 0.89(-0.08/0.05) &   & 0.71(-0.17/0.12) & 0.59(-0.21/0.15) \\
$c_{\mathrm{con}}$ & 0.90(-0.09/0.05) & 0.89(-0.11/0.06) &   & 0.90(-0.10/0.05) &   & 0.94(-0.06/0.03) & 0.89(-0.13/0.07) &   & 0.76(-0.07/0.03) & 0.62(-0.08/0.04) \\
$c_{\mathrm{er}}$ & 0.94(-0.09/0.04) & 0.92(-0.12/0.05) &   & 0.93(-0.11/0.04) &   & 0.95(-0.09/0.04) & 0.86(-0.16/0.07) &   & 0.93(-0.29/0.14) & 0.92(-0.39/0.22) \\
\enddata
~\\
\end{deluxetable*}

Figures \ref{fig:ar_b_goes}a and \ref{fig:ar_b_goes}b show the GOES SXR peak flux  $F_{\mathrm{SXR}}$ as a function of $A_{\mathrm{FL}}$ and  $\bar{\abs{B}}_{\mathrm{FL}}$, respectively, which also reveal distinct  correlations, but in case of $F_{\mathrm{SXR}}$ versus $\bar{\abs{B}}_{\mathrm{FL}}$ the correlation coefficient being significantly higher for eruptive ($c \approx 0.9$) than for confined ($c \approx 0.6$)  flares.
 For a certain mean magnetic flux density $\bar{\abs{B}}_{\mathrm{FL}}$ eruptive flares yield a higher $F_{\mathrm{SXR}}$ than confined flares. 
While there exists a distinct correlation between the GOES peak flux and the flare area, it is worth mentioning that eruptive flares of high GOES class seem to involve a similar range of areas, and the increase of $F_{\mathrm{SXR}}$ with the reconnection flux is mostly due to higher magnetic flux densities involved in these cases.   

In Table \ref{tab:fitcc}, we list the correlation coefficients and the fit parameters obtained from the regression  (linear model applied in log-log space) for the different relations plotted in Figures \ref{fig:phi_goes_sep}--\ref{fig:ar_b_goes}. We applied significance tests to all the correlations using Steiger's z test. All the correlations in Table \ref{tab:fitcc} are significant on the 99.99\% level (p=0.0001). In addition, we also list the 95\% confidence range on our correlation coefficients based on Fisher's transformation. 

Figure \ref{fig:dist_phi_ar} shows the flare reconnection flux $\varphi_{\mathrm{FL}}$ against the
total flux $\varphi_{\mathrm{AR}}$ of the source AR that produces the flare. The ARs in our study host magnetic fluxes between $1.9\cdot 10^{21}\, \mathrm{Mx}$ and $8.3\cdot 10^{22}\, \mathrm{Mx}$, 
whereas the flare reconnection fluxes range from $1.7\cdot 10^{20}\, \mathrm{Mx}$ to 
$2.5\cdot 10^{22}\, \mathrm{Mx}$.
From the figure it is seen that small events (i.e.\ small flare reconnection fluxes) can result from a large variety of AR fluxes, whereas the largest events require ARs with a high magnetic flux content.

Figure \ref{fig:ratio_goes} shows the ratio $r$ of the flare reconnection flux and the AR flux as a function of $F_{\mathrm{SXR}}$. We find a minimum of $r\approx0.03$, meaning that only 3\%\ of the AR magnetic flux is involved in the flare. In contrast, we find a maximum value of $r=0.46$, i.e. almost half of the AR flux was involved in the eruptive 3B/X3.8 flare on 2005 January 17 (event \#15). Figure \ref{fig:ratio_goes} shows that in small events (i.e. class B and C) the AR magnetic flux involved can range from a few percent up to 30\%, whereas in flares $\ge$M1 (except for the 2N/X2.0 flare on 2014 October 26) at least 10\% of the AR flux was involved. For flares $\gtrsim$X4, more than $\approx$30\%\ of the AR flux was involved in the flare reconnection. 
Note that in these largest events also the total flux contained in the source AR is considerably higher than in smaller flares (cf.\ Fig. \ref{fig:dist_phi_ar}).

Figure \ref{fig:cme_speed} shows the dependency of the CME speed on the flare reconnection flux. We find that for higher $\varphi_{\mathrm{FL}}$ the speed of the associated CME tends to be higher. The CMEs in our study have speeds ranging between $v_{\mathrm{CME}}=398\, \mathrm{km\,s^{-1}}$ and $v_{\mathrm{CME}}=2547\, \mathrm{km\,s^{-1}}$. We find a high correlation with a correlation coefficient of $c_{\mathrm{CME}}=0.84 \,(-0.22/+0.10)$ in linear space.

\section{Summary and discussion}
\label{sec:summary}

We analyzed magnetic reconnection rates and fluxes of a set of 51 flares, covering 19 eruptive and 32 confined events. Our main focus was to derive reconnection parameters over a large range in flare energy and to the study differences in confined and eruptive flares. In addition, we were interested whether the results obtained allow us to draw estimates on the largest flares that may occur on the Sun. Our study delivered the following main results:
\begin{enumerate}
 \item We find a very high correlation of the peak GOES SXR flux, $F_{\mathrm{SXR}}$, with the flare reconnection flux, $\varphi_{\mathrm{FL}}$, of $c=0.92 \left(-0.05/+0.03\right)$ in log-log space. 
 This value also implies that 85\% of the data can be explained by the regression model.
 There is no difference for confined or eruptive flares, both lie close to the same regression line. While $F_{\mathrm{SXR}}$ covers a range of over four orders of magnitude, we find a range of two
  orders of magnitude for $\varphi_{\mathrm{FL}}$.
 \item We find also a very high correlation between the peak of the GOES SXR time derivative $\dot{F}_{\mathrm{SXR}}$ and the peak reconnection rate $\dot\varphi_{\mathrm{FL}}$ with $c=0.90 \left(-0.10/+0.05\right)$. The fit parameters are significantly different between confined and eruptive flares.    
 \item We find that confined and eruptive flares significantly differ only regarding the ribbon area and the mean magnetic flux density swept by the ribbons. In the largest events, the flare areas reach values up to about $10^{10}\, \mathrm{km^2}$ and the mean photospheric magnetic flux density underlying the flare ribbons has values from $ \approx$100 G up to 800 G. For a given GOES class, the mean magnetic flux density in the flare ribbons is larger for confined than for eruptive events, on average by a factor of 2. Correspondingly, for a given reconnection flux we find smaller flare ribbon areas in confined flares. 
 These findings are consistent with the tendency of confined flares to occur closer to the flux-weighted center of ARs \citep[][Baumgartner et al., ApJ submitted]{Wang2007,Cheng2011}, where the mean magnetic flux density swept by the flare ribbons is expected to be larger. 
 \item We find that the fraction of the AR magnetic flux that is involved in the flare reconnection process  
ranges between 3\%\ and 46\%. It increases with the GOES class, amounting to at least 10\%\ of the AR magnetic flux in flares $\ge$ M1. For the largest flares ($\gtrsim$ X4), we find that at least 30\%\ of the AR flux is involved.    
 \item For eruptive flares, we find a high correlation of $c=0.84 \,(-0.22/+0.10)$ between the flare reconnection flux $\varphi_{\mathrm{FL}}$ and the speed of the associated CME, in agreement to the correlation reported by \cite{Qiu2005} of $c=0.89$. 
\end{enumerate}

In the following we place our main findings, as listed above, in the context of earlier studies within the research field, before we discuss their relevance in terms of the largest flares that may occur on the Sun. Our study reveals a higher correlation between $F_{\mathrm{SXR}}$ and $\varphi_{\mathrm{FL}}$ ($c = 0.92$), compared to the correlation found by \cite{Veronig2015} with $c=0.78$. The reason for the higher correlation presented here is most likely the use of a homogeneous data set and the application of the same method to all events in order to derive the relevant physical quantities, in contrast to that earlier study in which the results for 27 events from 4 different studies were combined. Recent statistical studies were published by \cite{Toriumi2017} and \cite{Kazachenko2017}. In contrast to these studies, which analyse flares \textgreater M5.0 and \textgreater C1.0, respectively, and the largest event included is of class X5.4,  we use a set of flare events covering a substantially wider range of flare classes, ranging from class B3.1 to X17. 
Another difference is that in our study we use H$\alpha$\ filtergrams, whereas \cite{Toriumi2017} and \cite{Kazachenko2017} use SDO AIA 1600 \AA\ images for the analysis of flare ribbons. \cite{Toriumi2017} found a correlation between $\varphi_{\mathrm{FL}}$ and $F_{\mathrm{SXR}}$ of $c=0.37$, while \cite{Kazachenko2017} found a Spearman ranking correlation coefficient of 0.66.  
The substantially higher correlation found in our study ($c\simeq 0.9$) is most likely due 
to the larger range of flares included. Our data set spans four orders of flare class, whereas e.g. the study of \cite{Toriumi2017} spans just one order of magnitude. If there is some intrinsic correlation between two data sets, which may be affected by measurement uncertainties and noise, then the larger the base range of the values included, the better this intrinsic correlation can be recovered from the data, resulting in a higher correlation coefficient.

\begin{figure}
 \includegraphics[width=\textwidth]{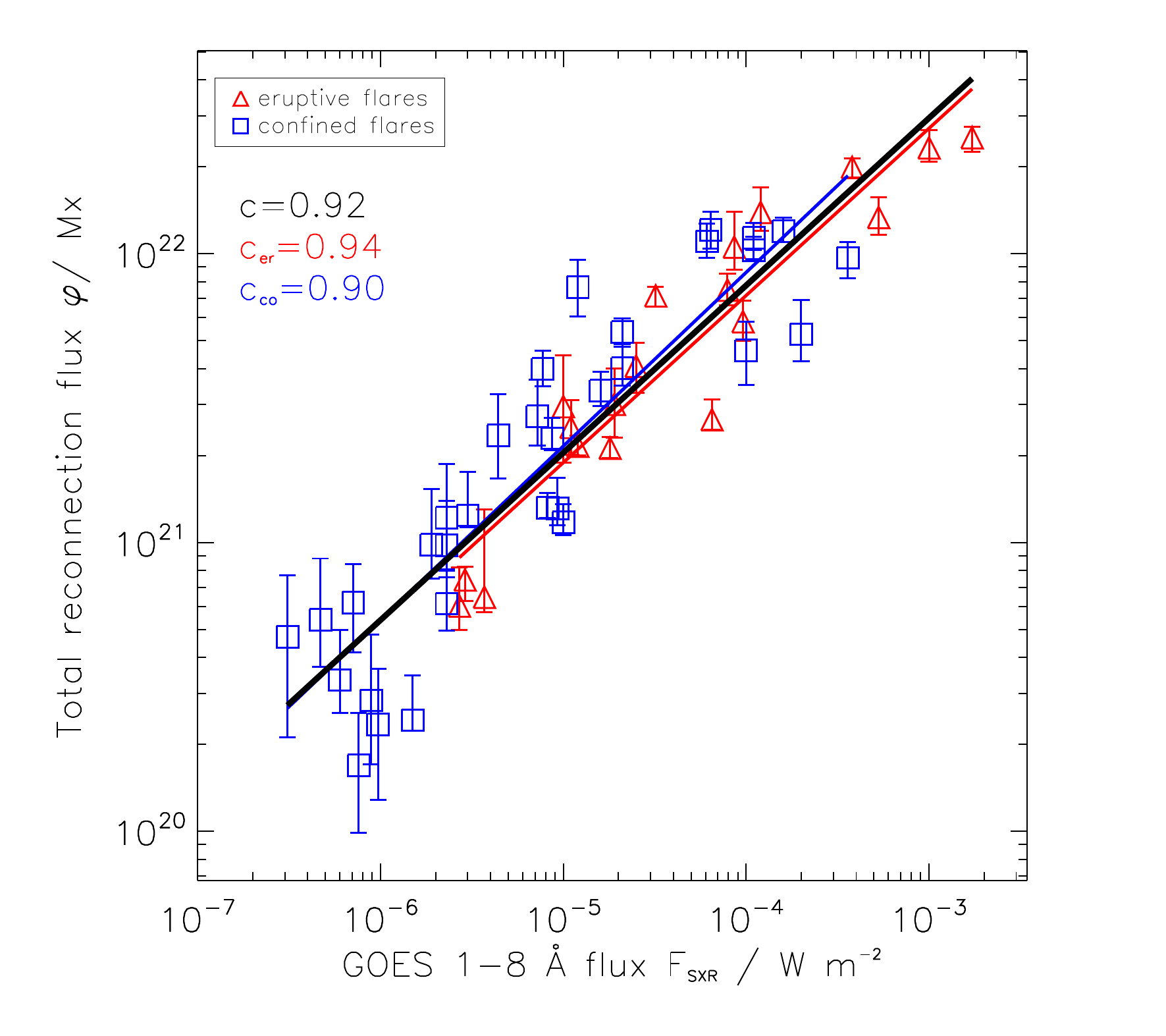}
 \caption{Total flare reconnection flux $\varphi_{\mathrm{FL}}$ versus GOES 1-8 \AA\ SXR peak flux $F_{\mathrm{SXR}}$. Blue squares indicate confined, red triangles eruptive events. The thick black lines (all events), thin red (eruptive events) and blue (confined events) lines represent the linear regression derived in log-log space.}
 \label{fig:phi_goes_sep}
\end{figure}

\begin{figure}
 \gridline{ \fig{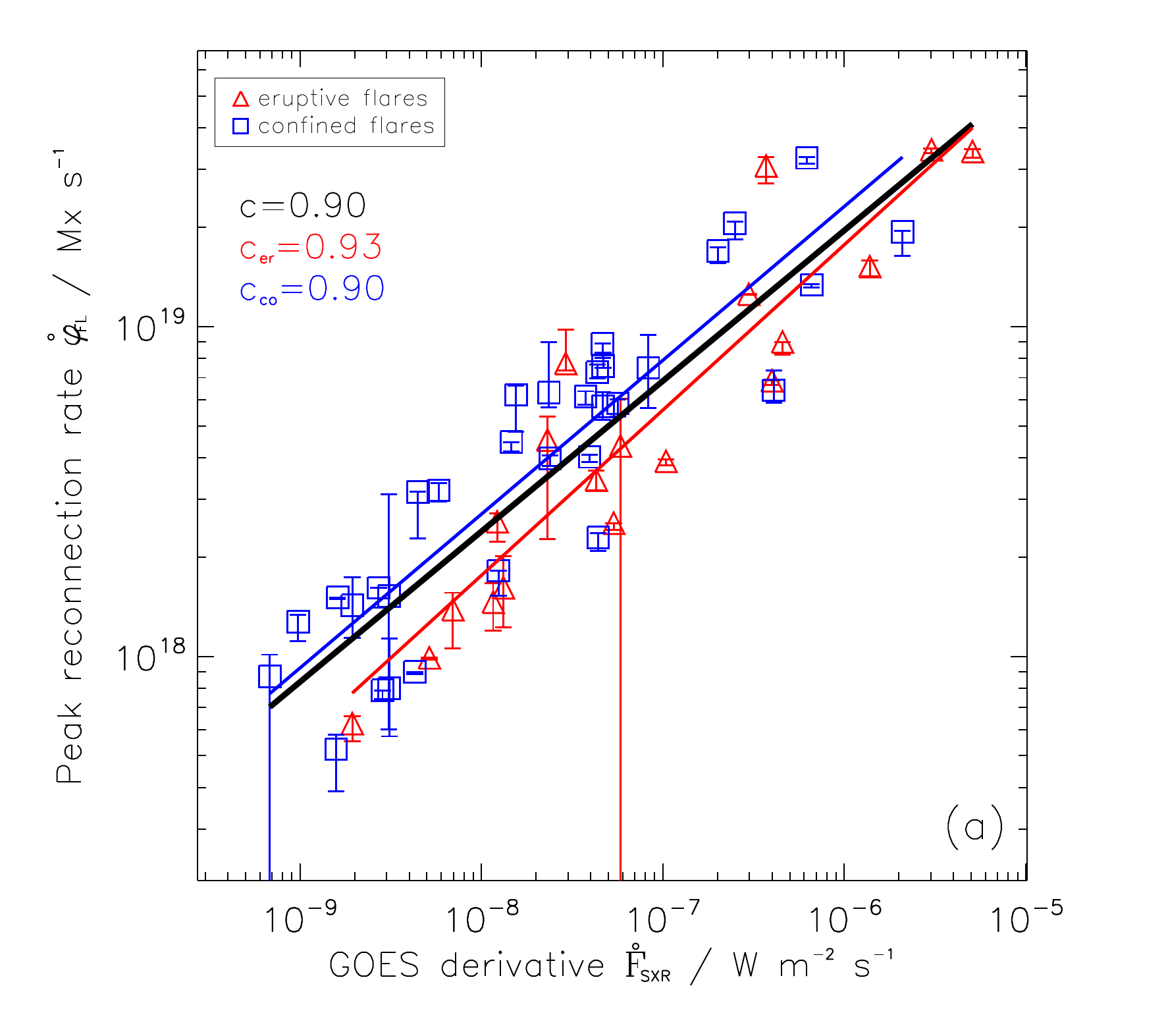}{0.6\textwidth}{\label{fig:dphi_goesderiv}}}
 \gridline{ \fig{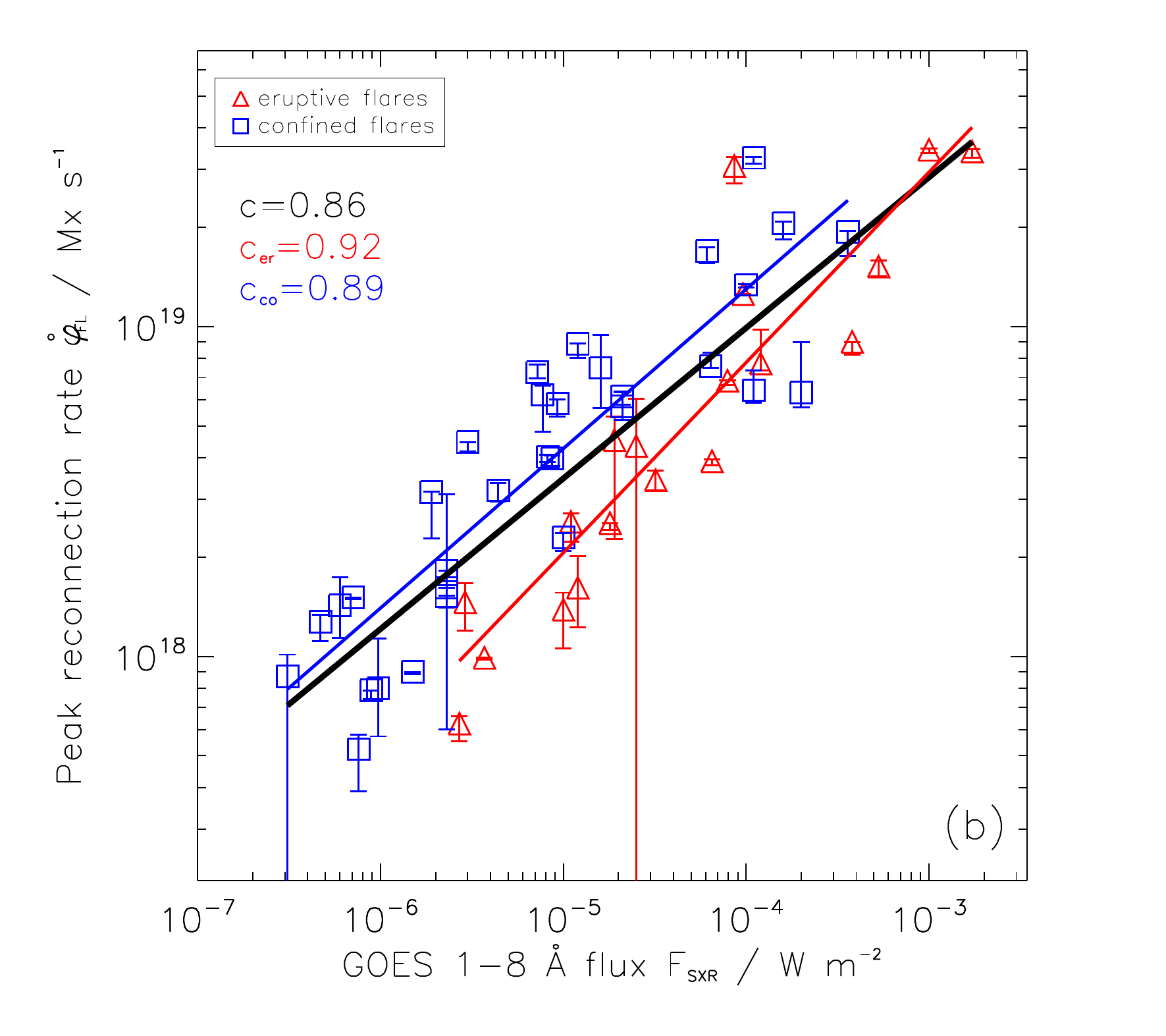}{0.6\textwidth}{\label{fig:dphi_goesflux}}}
  \caption{Peak of flare reconnection rate $\dot{\varphi}_{\mathrm{FL}}$ versus (a) the peak of the GOES 1-8 \AA\ SXR flux derivative $\dot{F}_{\mathrm{SXR}}$, (b) GOES 1-8 \AA\ peak flux $F_{\mathrm{SXR}}$. Blue squares indicate confined, red triangles eruptive events. The thick black lines (all events), thin red (eruptive events) and blue (confined events) lines represent the linear regression derived in log-log space.}
  \label{fig:scatter_plots}
\end{figure}

\begin{figure}
 \gridline{ \fig{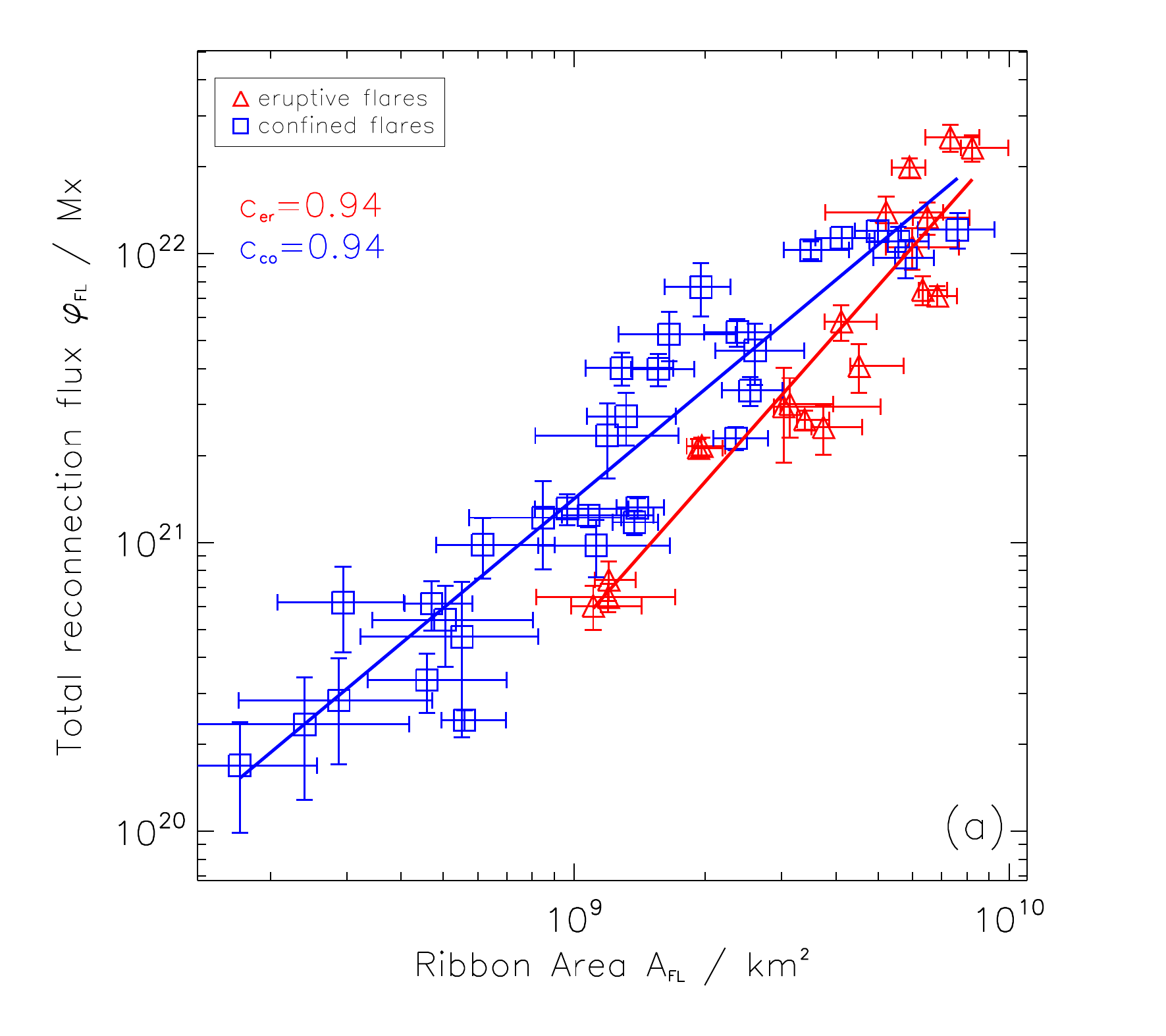}{0.6\textwidth}{}}
 \gridline{ \fig{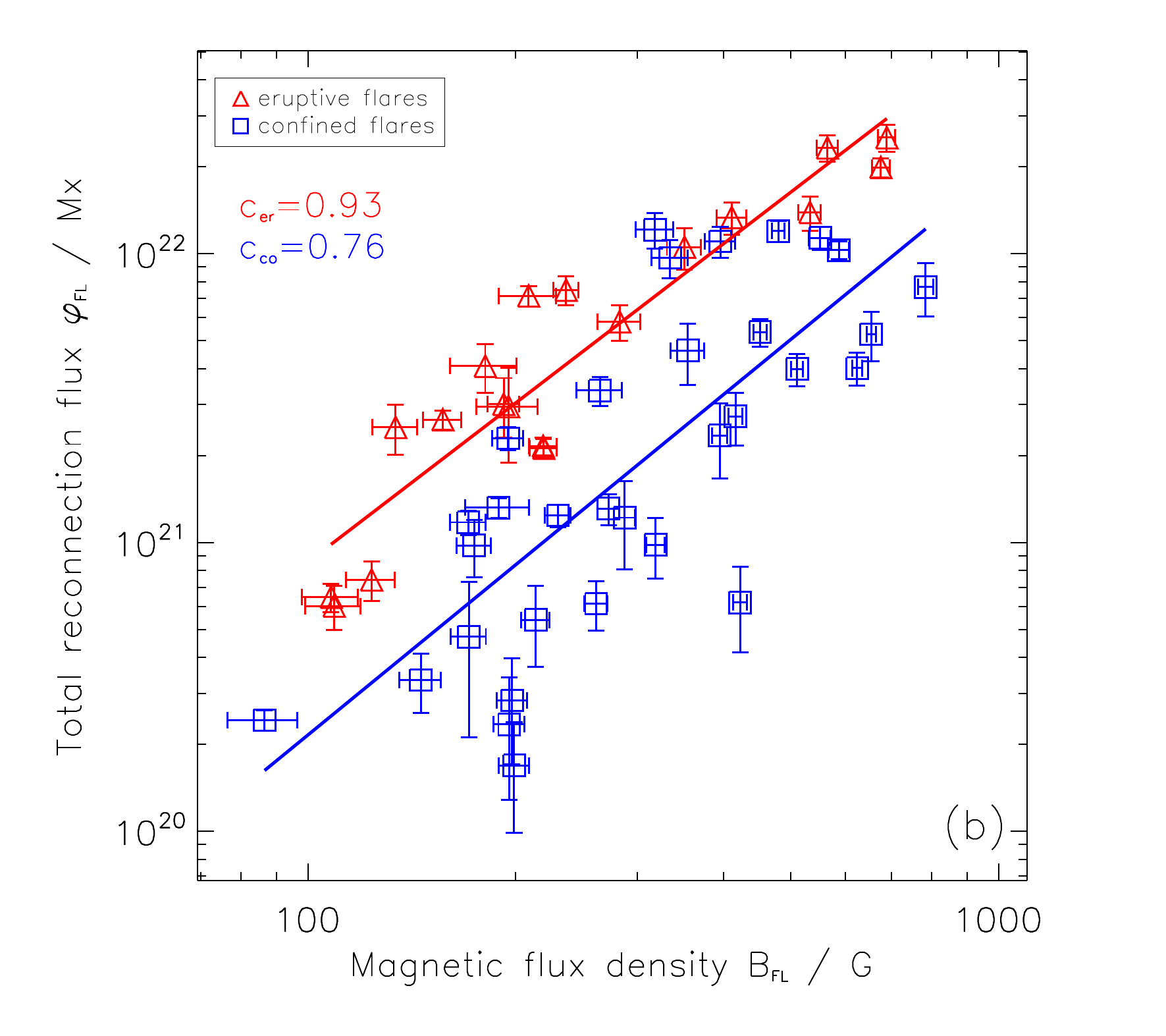}{0.6\textwidth}{}}
 \caption{Total flare reconnection flux $\varphi_{\mathrm{FL}}$ versus (a) flare ribbon area $A_{\mathrm{FL}}$, (b) mean magnetic flux density $\bar{\abs{B}}_{\mathrm{FL}}$ in the flare ribbons. Blue squares indicate confined, red triangles eruptive events.}
 \label{fig:ar_b_phi}
\end{figure}

\begin{figure}
 \gridline{ \fig{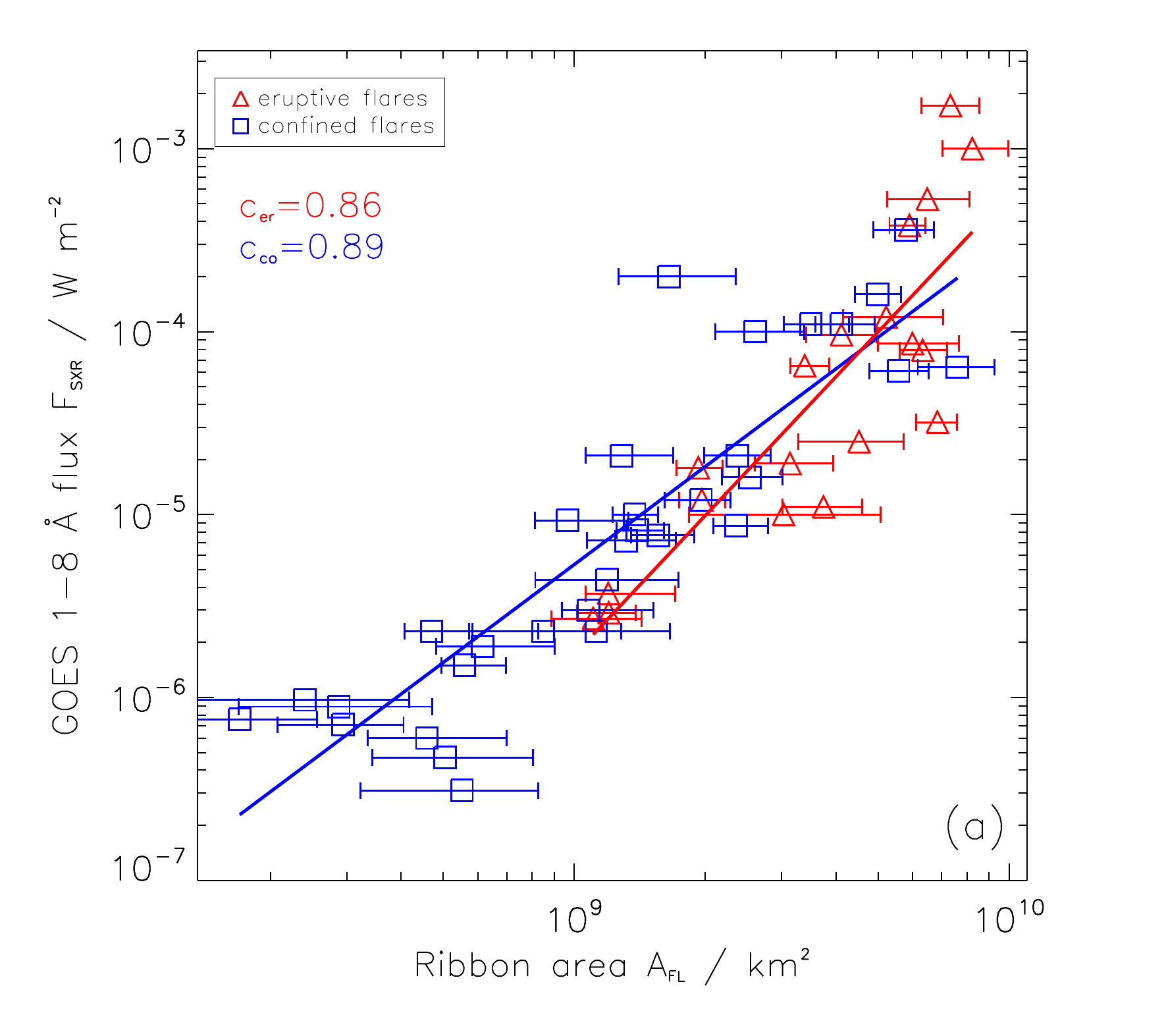}{0.6\textwidth}{}}
 \gridline{ \fig{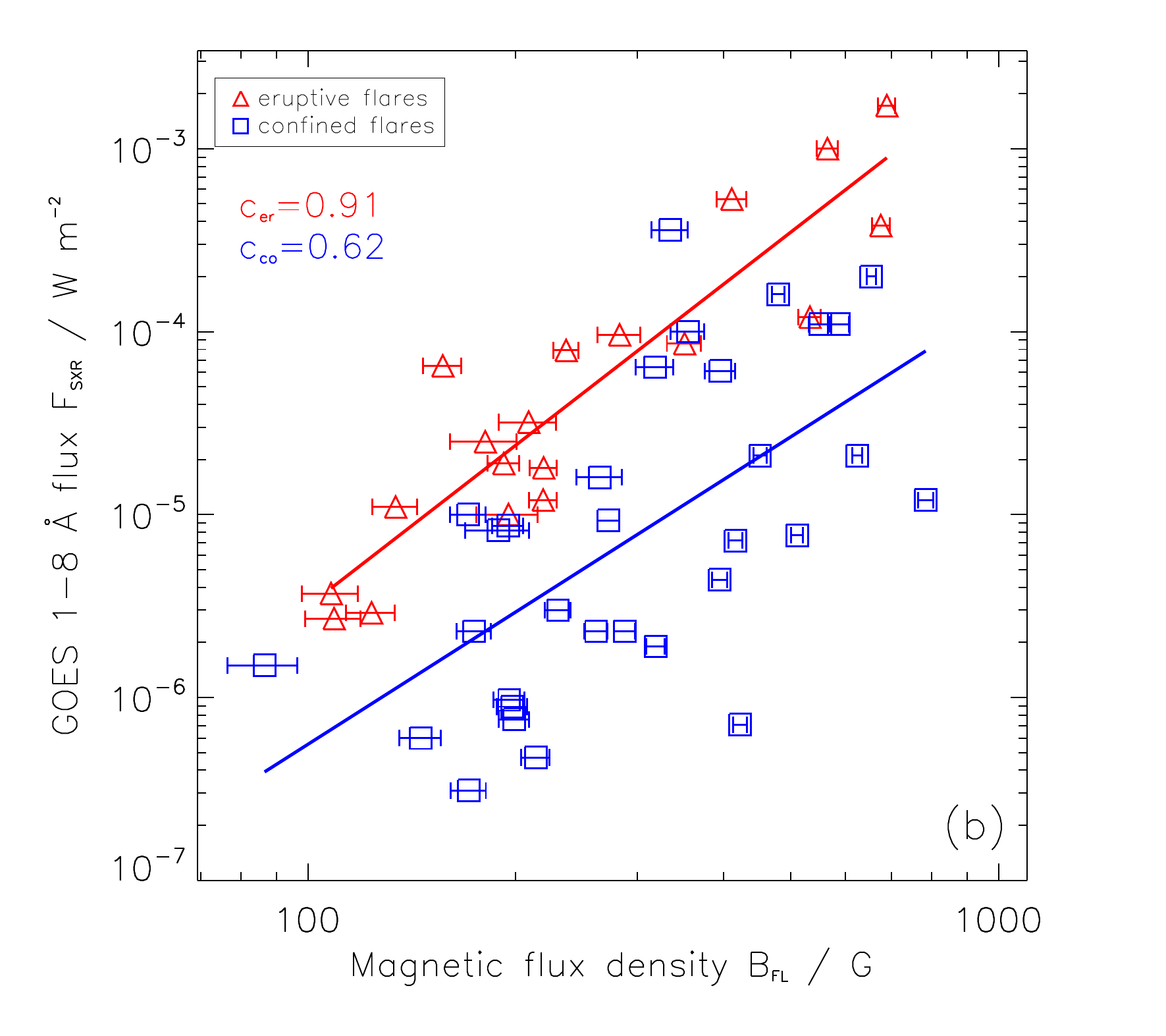}{0.6\textwidth}{}}
 \caption{(a) GOES soft X-ray peak flux $F_{\mathrm{SXR}}$ versus (a) flare ribbon area $A_{\mathrm{FL}}$, (b) mean magnetic flux density $\bar{\abs{B}}_{\mathrm{FL}}$ in the flare ribbons. Blue squares indicate confined, red triangles eruptive events.}
 \label{fig:ar_b_goes}
\end{figure}

\begin{figure}
 \begin{center}
  \includegraphics[width=\textwidth]{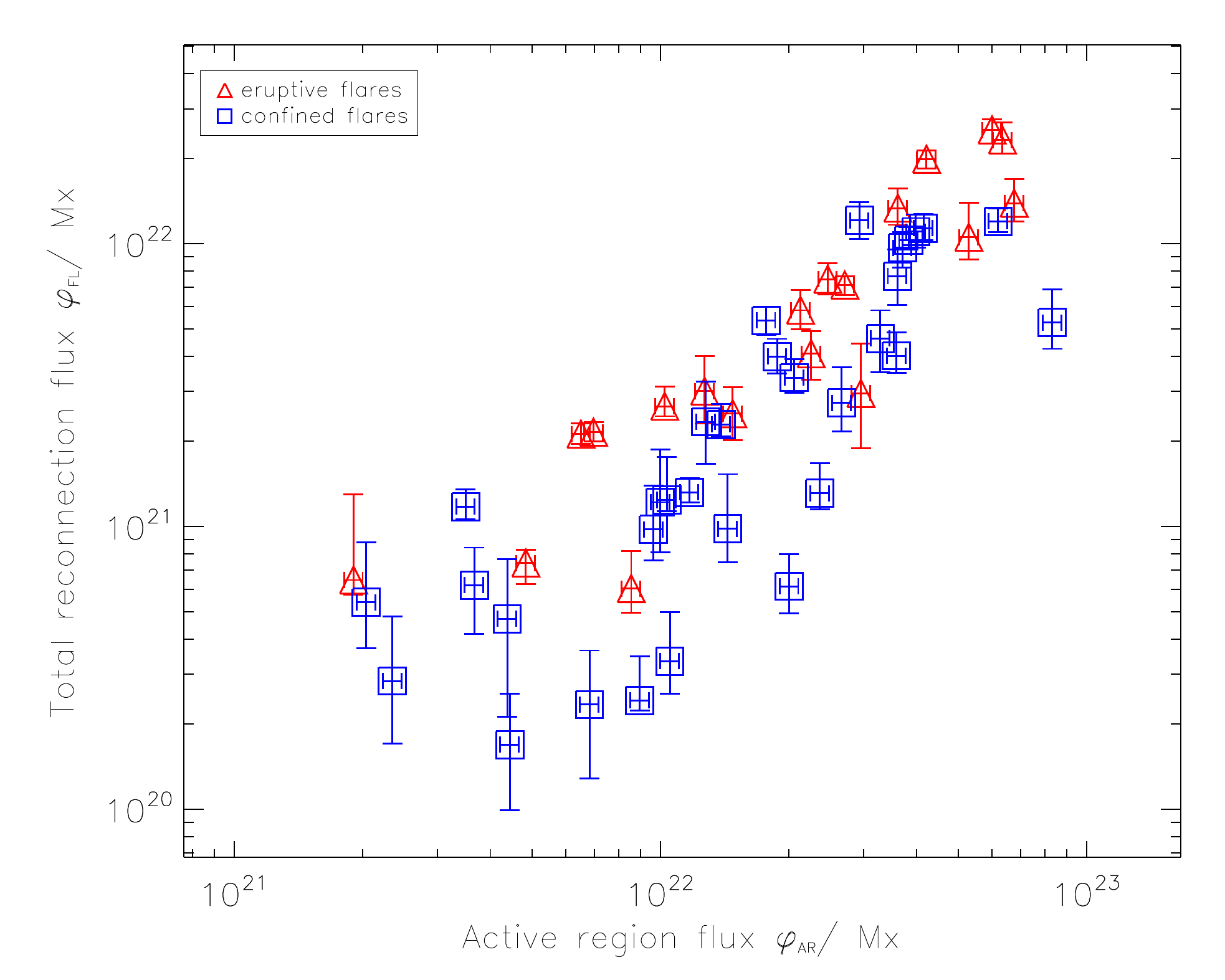}
  \caption{  Flare reconnection flux $\varphi_{\mathrm{FL}}$ against the total magnetic flux in the AR, $\varphi_{\mathrm{AR}}$, that causes the flare. Blue squares indicate confined, red triangles eruptive events.  }
  \label{fig:dist_phi_ar}
 \end{center}
\end{figure}

\begin{figure}
 \begin{center}
  \includegraphics[width=\textwidth]{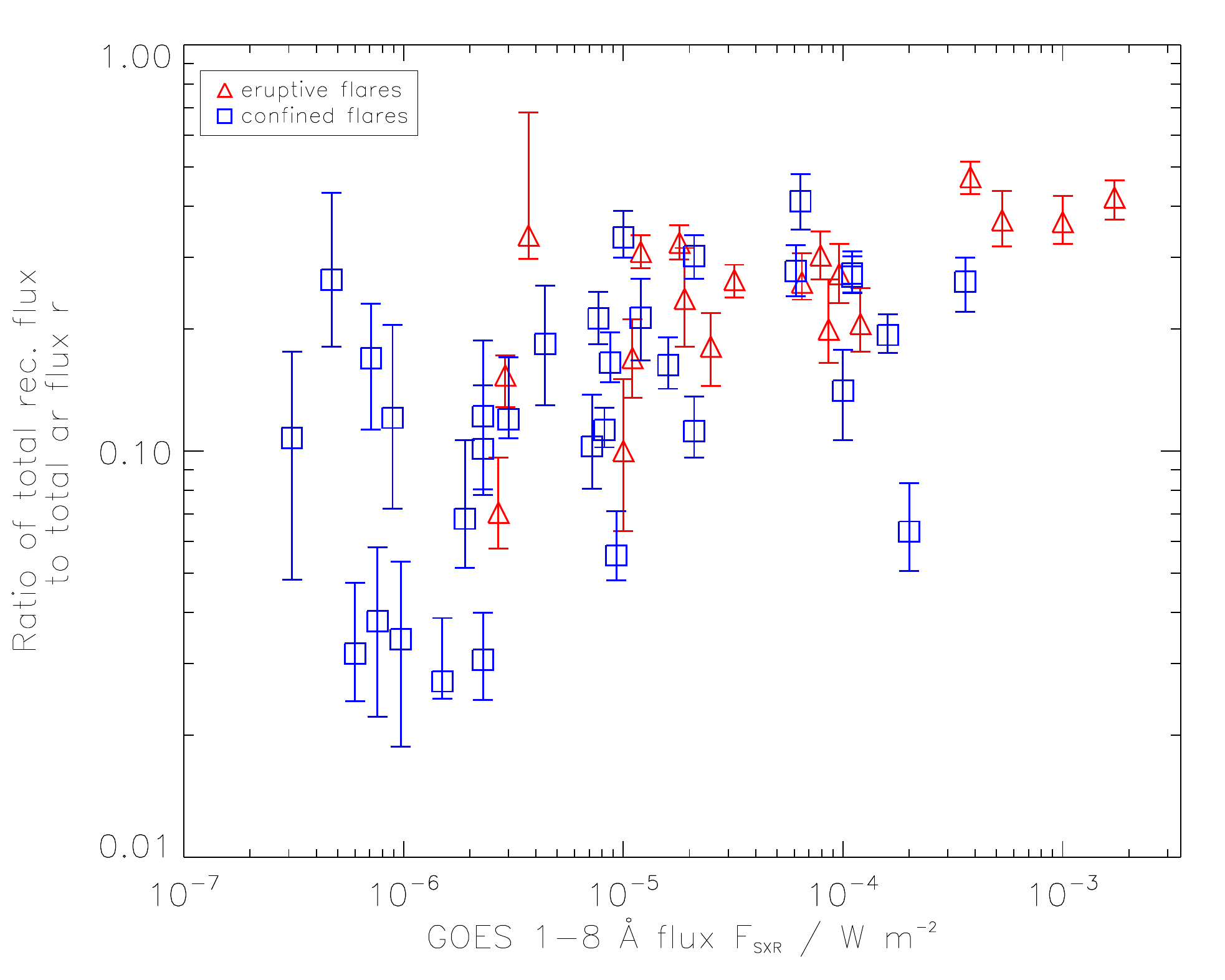}
  \caption{Fraction of the total AR flux that is involved in the reconnection, $r=\varphi_{\mathrm{FL}}/\varphi_{\mathrm{AR}}$, versus the GOES 1-8 \AA\ peak flux $F_{\mathrm{SXR}}$. Blue squares indicate confined, red triangles eruptive events.}
  \label{fig:ratio_goes}
 \end{center}
\end{figure}

\begin{figure}
 \begin{center}
  \includegraphics[width=\textwidth]{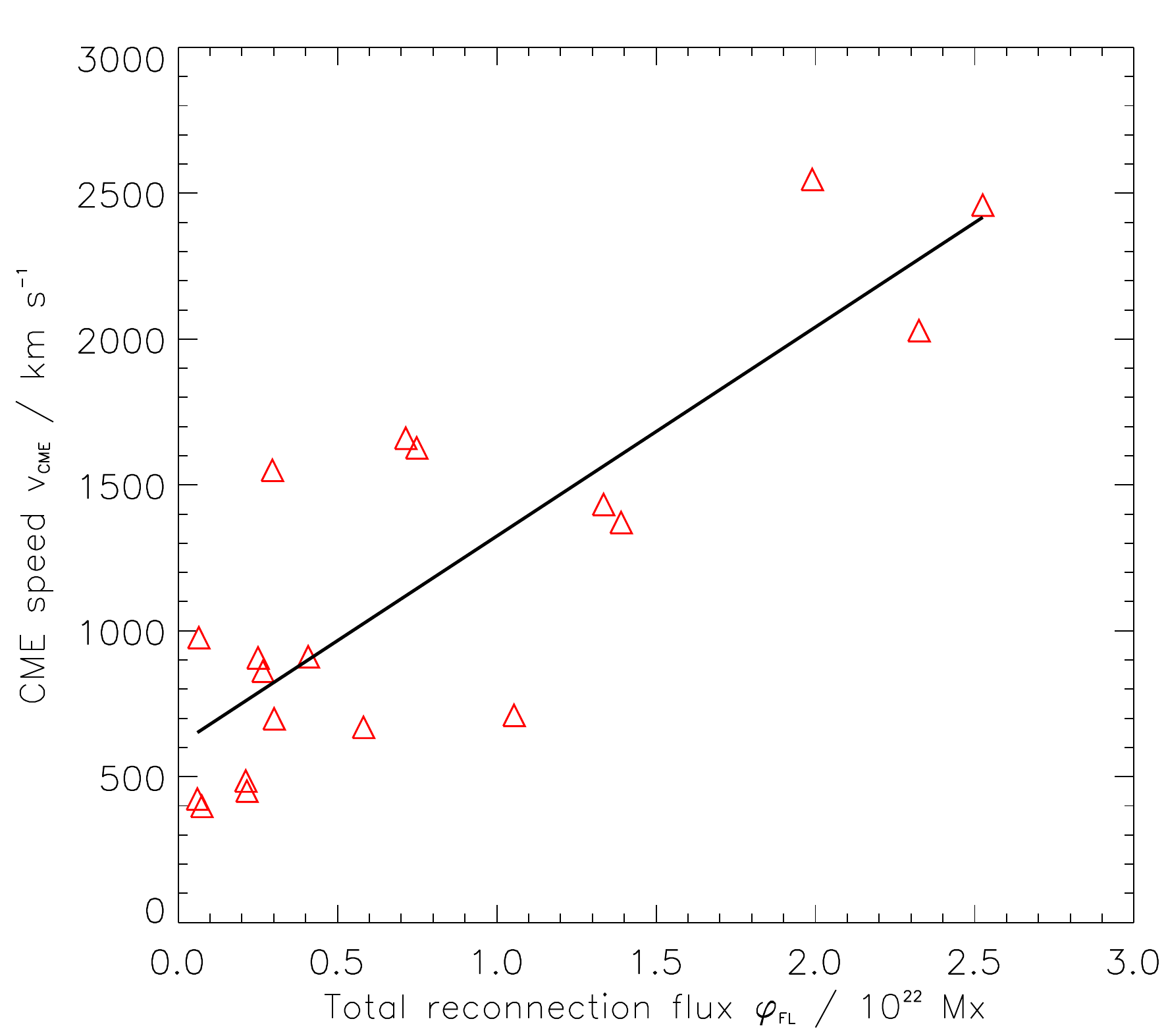}
  \caption{CME speed versus flare reconnection flux $\varphi_{\mathrm{FL}}$.}
  \label{fig:cme_speed}
 \end{center}
\end{figure}

\cite{Toriumi2017} found a linear regression between $F_{\mathrm{SXR}}$ and $\varphi_{\mathrm{FL}}$ less steep than we find in our study, which we also attribute to the different data range.
  \cite{Kazachenko2017} fitted a power law in the form of $y=d\,x^{a}$ to their data set, using $x = \varphi_{\mathrm{FL}}$ and  $y=F_{\mathrm{SXR}}$, finding $a= 1.53$. We find a similar power law index of $a=1.47 \pm 0.09$ when using the same notation (cf. Fig. \ref{fig:phi_goes_sep} and Table \ref{tab:fitcc}.). \cite{Kazachenko2017} speculated about the existence of a good correlation between the peak reconnection rate and the peak HXR flux. In our study, using the time derivative of the GOES SXR flux as an approximation for the HXR time profile, we were able to prove this strong dependency ($c\simeq 0.9$). 

Our results show a very tight correlation between the total flare reconnection flux and the GOES SXR peak flux, $c = 0.92$,  over four orders of magnitude in flare class  and two orders in the involved flare reconnection fluxes, with a typical scatter of about 0.3 orders of magnitude (cf. Fig. \ref{fig:phi_goes_sep}).
The linear regression model (in log-log space) can account for as much as 85\% of the variation of the data.  These findings allow us to quite accurately specify the relation between the two quantities, which in the following can also be used to extrapolate the relation towards the largest flares that may occur on the present day's Sun, depending on the total magnetic flux contained in the source AR. These considerations are interesting by their implications for solar activity, but in addition also very important in terms of estimating the strongest space weather events that may affect Earth.  
In addition, such extrapolation that is well grounded by observed solar quantities, also allows us to place the strongest flares expected from the present day's Sun in context to ``superflares'' on solar-like stars that have been recently discovered in Kepler data \citep{Maehara2012,Shibayama2013}. The estimates of the energies of these stellar superflares are up to $10^{35}$ - $10^{36}\,\mathrm{erg}$, making them a factor of 100 up to 10000 times larger than the biggest solar flares observed so far.

In the present study, we find that in the largest solar events observed, say GOES classes $\gtrsim$X4, between 30 to 50\%\ of the magnetic flux of the source AR is involved in the flare reconnection process. These numbers are in accordance with the recent studies by \cite{Kazachenko2017} and \cite{Toriumi2017}. The largest fluxes measured for ARs associated to major flare activity range from several times $10^{22}\,\mathrm{Mx}$ up to a few times $10^{23}\,\mathrm{Mx}$ \citep[e.g.,][]{Zhang2010,Chen2011,Yang2017}. For instance, AR 12192 (covered in our study) was the largest AR on the Sun since 24 years, with an area of 2750 $\mu$hem. It hosted a maximum photospheric vertical magnetic flux of $\sim2\cdot 10^{23}\,\mathrm{Mx}$ on 2014 October 27 \citep{Sun2015}, and was the source of a large number of confined X-class flares \citep[see, e.g.,][]{Thalmann2015}. This AR was even bigger than the famous AR 10486 that was the source of the strongest ``Halloween'' events, including the X17 and X10 flares of 2003 October 28 and 29 which are covered in our study. 
AR 10486 also produced the strongest SXR flare that was recorded in the GOES era, the X28+ event on 2003 November 4, during which the GOES fluxes were saturated for a few minutes. However, as this event occurred on the solar limb, magnetic reconnection fluxes cannot be derived for it. The X17 flare of 2003 October 28 (included in our study) occurred when AR 10486 was close to disk center. It was the 4\textsuperscript{th} strongest flare recorded by the GOES satellites \citep[see e.g.,][]{Tsurutani2005}.

In principle, of course, it is the free magnetic energy stored in the corona of magnetically complex ARs and released during a flare via magnetic reconnection that is the most relevant and direct physical quantity describing the process. However, to calculate the energy in flares and CMEs from observations is a difficult task, and the uncertainties are an order of magnitude \citep[e.g.,][]{Veronig2005,Emslie2005, Emslie2012}. Estimates of the magnetic energy of an AR and even more specifically the free magnetic energy available to power flare/CME events are not directly accessible, as we cannot reliably measure the coronal magnetic field. Therefore, such estimates are usually based on advanced three-dimensional coronal magnetic field models, using the vector magnetic field measured in the photosphere. However, the uncertainties of these estimates are again up to an order of magnitude, depending on the input data \citep[e.g.,][]{Thalmann2008}, model approach \citep[e.g.,][]{DeRosa2009} and possibly other factors \cite[e.g.,][]{DeRosa2015}.

The big advantage in our study is that we relate the flare energy release to the magnetic reconnection fluxes derived from \emph{direct} photospheric and chromospheric observations. In addition, the errors in the derived quantities are within $\approx$30\%\ \citep[see the present study and][]{Qiu2005}. The  distinct correlation obtained between the GOES SXR peak flux (a robust indicator of flare energy) and the flare reconnection flux over four orders of magnitude in GOES class, together with the finding that in the largest events up to $\approx 50$\%\ of the total AR flux is involved in the flare reconnection, allows us to estimate the size of the largest flares that may occur on the present-day Sun.

From the occurrence rates of stellar superflares observed on solar-like stars, \cite{Maehara2012} derive that on the Sun superflares with energies of $10^{34}\, \mathrm{erg}$ may occur once in 800 years, and flares with $10^{35}\,  \mathrm{erg}$ once in 5000 years. \cite{Shibata2013} estimated from order-of-magnitude considerations of solar dynamo theory, that ARs on the Sun hosting a magnetic flux of $2\cdot 10^{23}\, \mathrm{Mx}$ can be produced within one solar cycle period, and may be able to power flares with energies up to $10^{34}\,\mathrm{erg}$. Note that the value of $2\cdot 10^{23}\, \mathrm{Mx}$ is actually confirmed from observations of the largest and most active solar ARs \citep[e.g.,][]{Chen2011,Sun2015,Yang2017}. 
\cite{Shibata2013} also estimate that a superflare with an energy of $10^{35}\,\mathrm{erg}$ would require the hosting AR to carry a magnetic flux of some $10^{24}\, \mathrm{Mx}$, which would take about 40 years to be generated by the solar dynamo. 
\cite{Aulanier2013} derived the maximum energy that could be released by a solar flare using a dimensionless  MHD model, and scaling it by the size of the largest AR and the highest flux densities observed  in sunspots, assuming that in large flares 30\% of the AR flux are involved in the magnetic reconnection process. Their estimate is that a flare with an energy up to $6 \cdot 10^{33}$ erg could be produced, which they note as to be about 10 times larger in energy than the X17 event from 28 October 2003. 

In our study we found that in the largest solar flares up to 50\% of the AR magnetic flux is involved in the flare reconnection process.  In addition, we established a very tight correlation between the flare reconnection flux, $\varphi_{\rm FL}$, and the peak of the GOES 1--8 {\AA} SXR flux, $F_{\rm SXR}$, with $c=0.92$ (in log-log space). For the set of events under study, we find values of $a=0.580$ and $d=24.21$ for the regression line, $\log\left(\varphi_{\rm FL}\right)=a\,\log\left(F_{\rm SXR}\right)+d$ (see first column in Table \ref{tab:fitcc}; and Fig. \ref{fig:phi_goes_sep}). 
Now, we want to use these findings to make a maximum estimate, i.e. an estimate of the largest flares that might be produced by the present day's Sun, based on the extreme values determined from the range of observed data for flare reconnection and AR fluxes.

\begin{figure}
 \includegraphics[width=\textwidth]{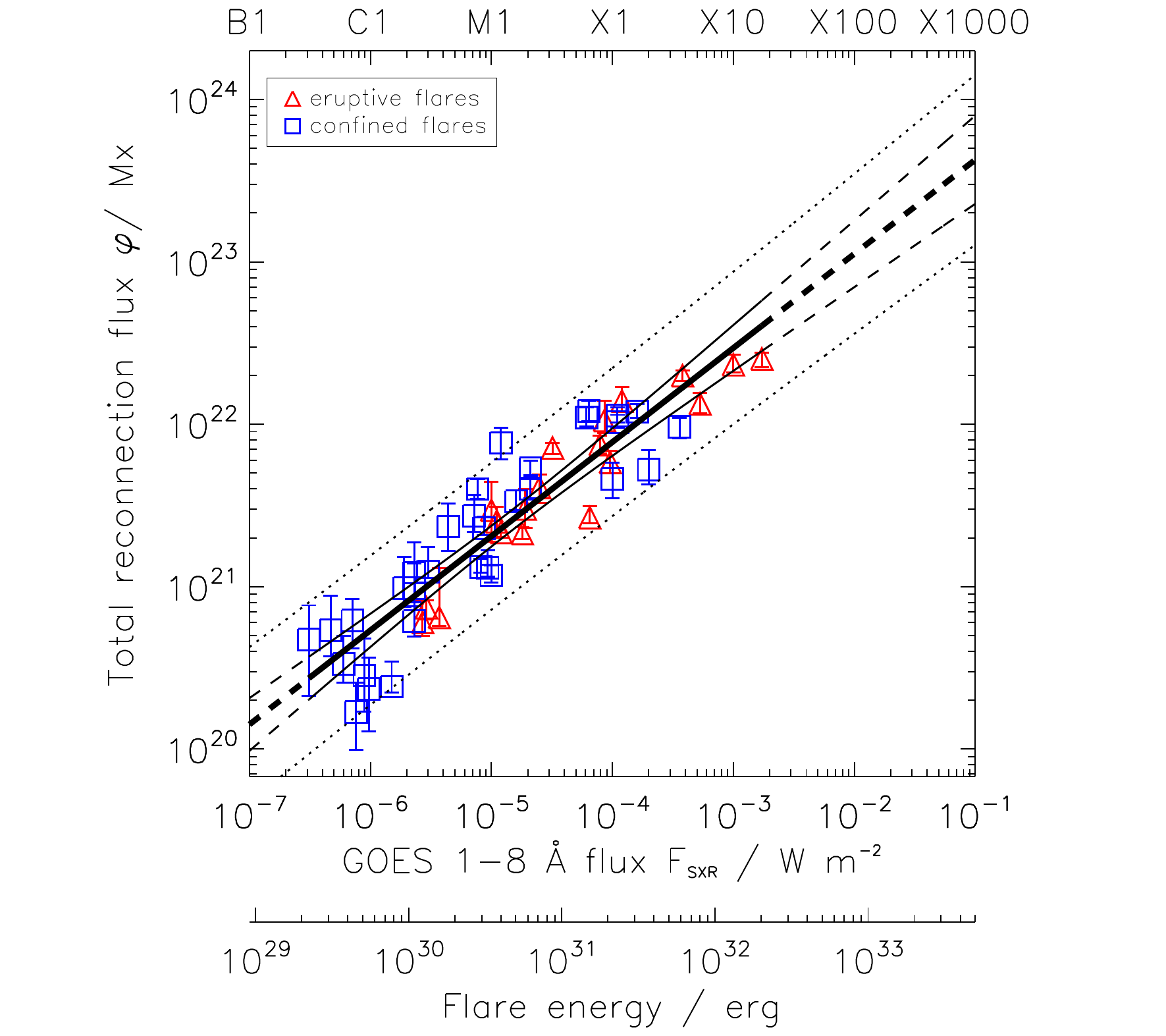}
 \caption{Total flare reconnection flux $\varphi_{\mathrm{FL}}$ versus GOES 1-8 \AA\ SXR peak flux $F_{\mathrm{SXR}}$. The second x-axis gives the corresponding bolometric flare energy using the relation derived in \cite{Kretzschmar2011}. Blue squares indicate confined, red triangles eruptive events. We plot the linear regression line derived in log-log space for all events (thick line) together with the 95\% confidence intervals (thin lines). Inside the measurement range, the regression line is drawn with solid lines, in the extrapolated ranges with dashed lines. The dotted curve outlines the prediction interval.
 }
 \label{fig:phi_goes_sep2}
\end{figure}

In Fig. \ref{fig:phi_goes_sep2} we re-plot this relation and the regression line for an extrapolation range up to energies corresponding to an X1000 flare. In addition, we plot the 95\% confidence interval and the prediction interval for individual data. Assuming the maximum percentage of  50\% of the magnetic flux contained in the AR contributing to the flare reconnection process,  we find that for an AR with a total flux of $2\cdot 10^{23}\,\mathrm{Mx}$  --- a value that is in accordance with the largest AR fluxes that have been measured  \citep{Zhang2010,Sun2015} --- a flare of GOES class X80 could be powered (with confidence bounds in the range X40 to X200). 
Note that in Fig. \ref{fig:phi_goes_sep2} we have also plotted a second x-axis, where we converted the GOES peak flux to bolometric flare energy. We use the scaling results from \cite{Kretzschmar2011}, who applied superposed epoch analysis on SOHO VIRGO sun-as-a-star measurements to derive the bolometric flare energy as function of GOES class. The relation is described by a power-law,
$\log\left(E_{\rm bol}\right)=a\,\log\left(F_{\rm SXR}\right)+d$  with $a=0.79\pm 0.10$ and $d=34.5\pm 0.5 $. Note that the extreme events studied in \cite{Emslie2012} also well follow that relation (cf.\ Fig.\ 8 in \cite{Warmuth2016}). Based on this relation, the bolometric energy of a flare of GOES class X80 is about $7 \cdot 10^{32}$ ergs.
Redoing the fit in Fig. 13 of reconnection fluxes as function of bolometric flare energy, we find 
$\log\left(\varphi_{\rm FL}\right)=a\,\log\left(E_{\rm bol}\right)+d$ 
with  $a=0.73\pm 0.04$ and $d=-1.1\pm 0.1 $. 

The largest AR ever observed on the Sun was in April 1947, with a size of approximately 6000 $\mu$hem \citep{Taylor1989}. \cite{Schrijver2012} estimated that the magnetic flux of this AR might have been as large as 
$6\cdot 10^{23}\,\mathrm{Mx}$. Such an AR could produce a flare of about class X500
(with confidence bounds in the range of X200 to X1000), corresponding to a bolometric energy of about $3 \cdot 10^{33}$ ergs. These estimates are about an order of magnitude larger than the largest flares that have been reported during the GOES era, and lie on the lower end of the energies of stellar superflares reported by \cite{Maehara2012}. Our data are in line with previous estimates from dynamo theory and MHD modeling \citep{Shibata2013,Aulanier2013}, and  indicate that the present day's Sun may be capable of producing a superflare and related space weather events that are at least an order of magnitude stronger than have been observed so far on the Sun.

\acknowledgments
This study was supported by the Austrian Science Fund (FWF): P27292-N20. We thank Dr. Bhuwan Joshi from the Physical Research Laboratory (PRL) for providing the USO H$\alpha$\ filtergrams and Dr. Chang Liu from the NJIT Space Weather Research Lab for the NSO H$\alpha$\ filtergrams. HMI data are courtesy of NASA/SDO and the HMI science teams. 

\bibliography{Rec_rates}

\begin{thebibliography}{}
\expandafter\ifx\csname natexlab\endcsname\relax\def\natexlab#1{#1}\fi
\providecommand{\url}[1]{\href{#1}{#1}}

\bibitem[{{Aulanier} {et~al.}(2013){Aulanier}, {D{\'e}moulin}, {Schrijver},
  {Janvier}, {Pariat}, \& {Schmieder}}]{Aulanier2013}
{Aulanier}, G., {D{\'e}moulin}, P., {Schrijver}, C.~J., {et~al.} 2013, \aap,
  549, A66

\bibitem[{{Benz}(2017)}]{Benz2017}
{Benz}, A.~O. 2017, Living Reviews in Solar Physics, 14, 2

\bibitem[{{Brown}(1971)}]{Brown1971}
{Brown}, J.~C. 1971, \solphys, 18, 489

\bibitem[{{Brown}(1973)}]{Brown1973}
---. 1973, \solphys, 31, 143

\bibitem[{{Carmichael}(1964)}]{Carmichael1964}
{Carmichael}, H. 1964, NASA Special Publication, 50, 451

\bibitem[{{Chen} {et~al.}(2011){Chen}, {Wang}, {Li}, {Feynman}, \&
  {Zhang}}]{Chen2011}
{Chen}, A.~Q., {Wang}, J.~X., {Li}, J.~W., {Feynman}, J., \& {Zhang}, J. 2011,
  \aap, 534, A47

\bibitem[{{Cheng} {et~al.}(2011){Cheng}, {Zhang}, {Ding}, {Guo}, \&
  {Su}}]{Cheng2011}
{Cheng}, X., {Zhang}, J., {Ding}, M.~D., {Guo}, Y., \& {Su}, J.~T. 2011, \apj,
  732, 87

\bibitem[{{Couvidat} {et~al.}(2016){Couvidat}, {Schou}, {Hoeksema}, {Bogart},
  {Bush}, {Duvall}, {Liu}, {Norton}, \& {Scherrer}}]{Couvidat2016}
{Couvidat}, S., {Schou}, J., {Hoeksema}, J.~T., {et~al.} 2016, \solphys, 291,
  1887

\bibitem[{{De Rosa} {et~al.}(2009){De Rosa}, {Schrijver}, {Barnes}, {Leka},
  {Lites}, {Aschwanden}, {Amari}, {Canou}, {McTiernan}, {R{\'e}gnier},
  {Thalmann}, {Valori}, {Wheatland}, {Wiegelmann}, {Cheung}, {Conlon},
  {Fuhrmann}, {Inhester}, \& {Tadesse}}]{DeRosa2009}
{De Rosa}, M.~L., {Schrijver}, C.~J., {Barnes}, G., {et~al.} 2009, \apj, 696,
  1780

\bibitem[{{Dennis} {et~al.}(2003){Dennis}, {Veronig}, {Schwartz}, {Sui},
  {Tolbert}, {Zarro}, \& {Rhessi Team}}]{Dennis2003}
{Dennis}, B.~R., {Veronig}, A., {Schwartz}, R.~A., {et~al.} 2003, Advances in
  Space Research, 32, 2459

\bibitem[{{Dennis} \& {Zarro}(1993)}]{Dennis1993}
{Dennis}, B.~R., \& {Zarro}, D.~M. 1993, \solphys, 146, 177

\bibitem[{{DeRosa} {et~al.}(2015){DeRosa}, {Wheatland}, {Leka}, {Barnes},
  {Amari}, {Canou}, {Gilchrist}, {Thalmann}, {Valori}, {Wiegelmann},
  {Schrijver}, {Malanushenko}, {Sun}, \& {R{\'e}gnier}}]{DeRosa2015}
{DeRosa}, M.~L., {Wheatland}, M.~S., {Leka}, K.~D., {et~al.} 2015, \apj, 811,
  107

\bibitem[{{Emslie} {et~al.}(2005){Emslie}, {Dennis}, {Holman}, \&
  {Hudson}}]{Emslie2005}
{Emslie}, A.~G., {Dennis}, B.~R., {Holman}, G.~D., \& {Hudson}, H.~S. 2005,
  Journal of Geophysical Research (Space Physics), 110, A11103

\bibitem[{{Emslie} {et~al.}(2012){Emslie}, {Dennis}, {Shih}, {Chamberlin},
  {Mewaldt}, {Moore}, {Share}, {Vourlidas}, \& {Welsch}}]{Emslie2012}
{Emslie}, A.~G., {Dennis}, B.~R., {Shih}, A.~Y., {et~al.} 2012, \apj, 759, 71

\bibitem[{{Fisher} {et~al.}(1985){Fisher}, {Canfield}, \&
  {McClymont}}]{Fisher1985}
{Fisher}, G.~H., {Canfield}, R.~C., \& {McClymont}, A.~N. 1985, \apj, 289, 414

\bibitem[{{Fletcher} {et~al.}(2011){Fletcher}, {Dennis}, {Hudson}, {Krucker},
  {Phillips}, {Veronig}, {Battaglia}, {Bone}, {Caspi}, {Chen}, {Gallagher},
  {Grigis}, {Ji}, {Liu}, {Milligan}, \& {Temmer}}]{Fletcher2011}
{Fletcher}, L., {Dennis}, B.~R., {Hudson}, H.~S., {et~al.} 2011, \ssr, 159, 19

\bibitem[{{Forbes} \& {Lin}(2000)}]{Forbes2000}
{Forbes}, T.~G., \& {Lin}, J. 2000, Journal of Atmospheric and
  Solar-Terrestrial Physics, 62, 1499

\bibitem[{{Forbes} \& {Priest}(1984)}]{Forbes1984}
{Forbes}, T.~G., \& {Priest}, E.~R. 1984, in Solar terrestrial physics: present
  and future, ed. D.~M. {Butler} \& K.~{Papadopoulos}, 1--35

\bibitem[{{Hirayama}(1974)}]{Hirayama1974}
{Hirayama}, T. 1974, \solphys, 34, 323

\bibitem[{{Holman}(2016)}]{Holman2016}
{Holman}, G.~D. 2016, Journal of Geophysical Research (Space Physics), 121, 11

\bibitem[{{Hudson}(1991)}]{Hudson1991a}
{Hudson}, H.~S. 1991, \solphys, 133, 357

\bibitem[{{Janvier} {et~al.}(2015){Janvier}, {Aulanier}, \&
  {D{\'e}moulin}}]{Janvier2015}
{Janvier}, M., {Aulanier}, G., \& {D{\'e}moulin}, P. 2015, \solphys, 290, 3425

\bibitem[{{Kazachenko} {et~al.}(2017){Kazachenko}, {Lynch}, {Welsch}, \&
  {Sun}}]{Kazachenko2017}
{Kazachenko}, M.~D., {Lynch}, B.~J., {Welsch}, B.~T., \& {Sun}, X. 2017, \apj,
  845, 49

\bibitem[{{Kopp} \& {Pneuman}(1976)}]{Kopp1976}
{Kopp}, R.~A., \& {Pneuman}, G.~W. 1976, \solphys, 50, 85

\bibitem[{{Kretzschmar}(2011)}]{Kretzschmar2011}
{Kretzschmar}, M. 2011, \aap, 530, A84

\bibitem[{{Maehara} {et~al.}(2012){Maehara}, {Shibayama}, {Notsu}, {Notsu},
  {Nagao}, {Kusaba}, {Honda}, {Nogami}, \& {Shibata}}]{Maehara2012}
{Maehara}, H., {Shibayama}, T., {Notsu}, S., {et~al.} 2012, \nat, 485, 478

\bibitem[{{Miklenic} {et~al.}(2009){Miklenic}, {Veronig}, \& {Vr{\v
  s}nak}}]{Miklenic2009}
{Miklenic}, C.~H., {Veronig}, A.~M., \& {Vr{\v s}nak}, B. 2009, \aap, 499, 893

\bibitem[{{Neidig} {et~al.}(1998){Neidig}, {Wiborg}, {Confer}, {Haas}, {Dunn},
  {Balasubramaniam}, {Gullixson}, {Craig}, {Kaufman}, {Hull}, {McGraw},
  {Henry}, {Rentschler}, {Keller}, {Jones}, {Coulter}, {Gregory}, {Schimming},
  \& {Smaga}}]{Neidig1998}
{Neidig}, D., {Wiborg}, P., {Confer}, M., {et~al.} 1998, in Astronomical
  Society of the Pacific Conference Series, Vol. 140, Synoptic Solar Physics,
  ed. K.~S. {Balasubramaniam}, J.~{Harvey}, \& D.~{Rabin}, 519

\bibitem[{{Neupert}(1968)}]{Neupert1968}
{Neupert}, W.~M. 1968, \apjl, 153, L59

\bibitem[{{Otruba}(1999)}]{Otruba1999}
{Otruba}, W. 1999, in Astronomical Society of the Pacific Conference Series,
  Vol. 184, Third Advances in Solar Physics Euroconference: Magnetic Fields and
  Oscillations, ed. B.~{Schmieder}, A.~{Hofmann}, \& J.~{Staude}, 314--318

\bibitem[{{Otruba} \& {P{\"o}tzi}(2003)}]{Otruba2003}
{Otruba}, W., \& {P{\"o}tzi}, W. 2003, Hvar Observatory Bulletin, 27, 189

\bibitem[{{P{\"o}tzi} {et~al.}(2015){P{\"o}tzi}, {Veronig}, {Riegler},
  {Amerstorfer}, {Pock}, {Temmer}, {Polanec}, \& {Baumgartner}}]{Poetzi2015}
{P{\"o}tzi}, W., {Veronig}, A.~M., {Riegler}, G., {et~al.} 2015, \solphys, 290,
  951

\bibitem[{{Priest} \& {Forbes}(2002)}]{Priest2002}
{Priest}, E.~R., \& {Forbes}, T.~G. 2002, \aapr, 10, 313

\bibitem[{{Qiu} \& {Yurchyshyn}(2005)}]{Qiu2005}
{Qiu}, J., \& {Yurchyshyn}, V.~B. 2005, \apjl, 634, L121

\bibitem[{{Scherrer} {et~al.}(1995){Scherrer}, {Bogart}, {Bush}, {Hoeksema},
  {Kosovichev}, {Schou}, {Rosenberg}, {Springer}, {Tarbell}, {Title},
  {Wolfson}, {Zayer}, \& {MDI Engineering Team}}]{Scherrer1995}
{Scherrer}, P.~H., {Bogart}, R.~S., {Bush}, R.~I., {et~al.} 1995, \solphys,
  162, 129

\bibitem[{{Schou} {et~al.}(2012){Schou}, {Scherrer}, {Bush}, {Wachter},
  {Couvidat}, {Rabello-Soares}, {Bogart}, {Hoeksema}, {Liu}, {Duvall}, {Akin},
  {Allard}, {Miles}, {Rairden}, {Shine}, {Tarbell}, {Title}, {Wolfson},
  {Elmore}, {Norton}, \& {Tomczyk}}]{Schou2012}
{Schou}, J., {Scherrer}, P.~H., {Bush}, R.~I., {et~al.} 2012, \solphys, 275,
  229

\bibitem[{{Schrijver} {et~al.}(2012){Schrijver}, {Beer}, {Baltensperger},
  {Cliver}, {G{\"u}del}, {Hudson}, {McCracken}, {Osten}, {Peter}, {Soderblom},
  {Usoskin}, \& {Wolff}}]{Schrijver2012}
{Schrijver}, C.~J., {Beer}, J., {Baltensperger}, U., {et~al.} 2012, Journal of
  Geophysical Research (Space Physics), 117, A08103

\bibitem[{{Shibata} \& {Magara}(2011)}]{Shibata2011}
{Shibata}, K., \& {Magara}, T. 2011, Living Reviews in Solar Physics, 8, 6

\bibitem[{{Shibata} {et~al.}(2013){Shibata}, {Isobe}, {Hillier}, {Choudhuri},
  {Maehara}, {Ishii}, {Shibayama}, {Notsu}, {Notsu}, {Nagao}, {Honda}, \&
  {Nogami}}]{Shibata2013}
{Shibata}, K., {Isobe}, H., {Hillier}, A., {et~al.} 2013, \pasj, 65, 49

\bibitem[{{Shibayama} {et~al.}(2013){Shibayama}, {Maehara}, {Notsu}, {Notsu},
  {Nagao}, {Honda}, {Ishii}, {Nogami}, \& {Shibata}}]{Shibayama2013}
{Shibayama}, T., {Maehara}, H., {Notsu}, S., {et~al.} 2013, \apjs, 209, 5

\bibitem[{{Sturrock}(1966)}]{Sturrock1966}
{Sturrock}, P.~A. 1966, \nat, 211, 695

\bibitem[{{Sun} {et~al.}(2015){Sun}, {Bobra}, {Hoeksema}, {Liu}, {Li}, {Shen},
  {Couvidat}, {Norton}, \& {Fisher}}]{Sun2015}
{Sun}, X., {Bobra}, M.~G., {Hoeksema}, J.~T., {et~al.} 2015, \apjl, 804, L28

\bibitem[{{Taylor}(1989)}]{Taylor1989}
{Taylor}, P.~O. 1989, Journal of the American Association of Variable Star
  Observers (JAAVSO), 18, 65

\bibitem[{{Temmer} {et~al.}(2017){Temmer}, {Thalmann}, {Dissauer}, {Veronig},
  {Tschernitz}, {Hinterreiter}, \& {Rodriguez}}]{Temmer2017}
{Temmer}, M., {Thalmann}, J.~K., {Dissauer}, K., {et~al.} 2017, \solphys, 292,
  93

\bibitem[{{Thalmann} {et~al.}(2015){Thalmann}, {Su}, {Temmer}, \&
  {Veronig}}]{Thalmann2015}
{Thalmann}, J.~K., {Su}, Y., {Temmer}, M., \& {Veronig}, A.~M. 2015, \apjl,
  801, L23

\bibitem[{{Thalmann} {et~al.}(2008){Thalmann}, {Wiegelmann}, \&
  {Raouafi}}]{Thalmann2008}
{Thalmann}, J.~K., {Wiegelmann}, T., \& {Raouafi}, N.-E. 2008, \aap, 488, L71

\bibitem[{{Toriumi} {et~al.}(2017){Toriumi}, {Schrijver}, {Harra}, {Hudson}, \&
  {Nagashima}}]{Toriumi2017}
{Toriumi}, S., {Schrijver}, C.~J., {Harra}, L.~K., {Hudson}, H., \&
  {Nagashima}, K. 2017, \apj, 834, 56

\bibitem[{{Tsurutani} {et~al.}(2005){Tsurutani}, {Judge}, {Guarnieri},
  {Gangopadhyay}, {Jones}, {Nuttall}, {Zambon}, {Didkovsky}, {Mannucci},
  {Iijima}, {Meier}, {Immel}, {Woods}, {Prasad}, {Floyd}, {Huba}, {Solomon},
  {Straus}, \& {Viereck}}]{Tsurutani2005}
{Tsurutani}, B.~T., {Judge}, D.~L., {Guarnieri}, F.~L., {et~al.} 2005, \grl,
  32, L03S09

\bibitem[{{Veronig} {et~al.}(2002){Veronig}, {Temmer}, {Hanslmeier}, {Otruba},
  \& {Messerotti}}]{Veronig2002}
{Veronig}, A., {Temmer}, M., {Hanslmeier}, A., {Otruba}, W., \& {Messerotti},
  M. 2002, \aap, 382, 1070

\bibitem[{{Veronig} {et~al.}(2005){Veronig}, {Brown}, {Dennis}, {Schwartz},
  {Sui}, \& {Tolbert}}]{Veronig2005}
{Veronig}, A.~M., {Brown}, J.~C., {Dennis}, B.~R., {et~al.} 2005, \apj, 621,
  482

\bibitem[{{Veronig} \& {Polanec}(2015)}]{Veronig2015}
{Veronig}, A.~M., \& {Polanec}, W. 2015, \solphys, 290, 2923

\bibitem[{{Wang} \& {Zhang}(2007)}]{Wang2007}
{Wang}, Y., \& {Zhang}, J. 2007, \apj, 665, 1428

\bibitem[{{Warmuth} \& {Mann}(2016)}]{Warmuth2016}
{Warmuth}, A., \& {Mann}, G. 2016, \aap, 588, A116

\bibitem[{{Yang} {et~al.}(2017){Yang}, {Hsieh}, {Yu}, \& {Chen}}]{Yang2017}
{Yang}, Y.-H., {Hsieh}, M.-S., {Yu}, H.-S., \& {Chen}, P.~F. 2017, \apj, 834,
  150

\bibitem[{{Yashiro} {et~al.}(2005){Yashiro}, {Gopalswamy}, {Akiyama},
  {Michalek}, \& {Howard}}]{Yashiro2005}
{Yashiro}, S., {Gopalswamy}, N., {Akiyama}, S., {Michalek}, G., \& {Howard},
  R.~A. 2005, Journal of Geophysical Research (Space Physics), 110, A12S05

\bibitem[{{Zhang} {et~al.}(2010){Zhang}, {Wang}, \& {Liu}}]{Zhang2010}
{Zhang}, J., {Wang}, Y., \& {Liu}, Y. 2010, \apj, 723, 1006

\end{thebibliography}
\bibliographystyle{aasjournal}
\clearpage

\end{document}